\documentclass[twocolumn,secnumarabic,amsmath,amssymb,nobibnotes,aps,pra,showkeys,superscriptaddress,floatfix,tightenlines]{revtex4-2}

\usepackage{amsfonts}
\usepackage[caption=false, font=footnotesize]{subfig}
\usepackage{graphicx}
\usepackage{wrapfig}
\usepackage[mathscr]{eucal}
\usepackage{soul}
\usepackage{tabularx}
\usepackage{makecell}
\usepackage{enumitem}

\usepackage[rawfloats=true]{floatrow} 
\restylefloat{figure}

\usepackage[colorlinks,citecolor=blue,urlcolor=blue,bookmarks=false,hypertexnames=true]{hyperref}

\begin{document}
	
	\title{Finding solutions to the integer case constraint satisfiability problem using Grover's algorithm}
	
	\author{Gayathree M.~Vinod, Anil Shaji}
	\email[Corresponding author: ]{gaya3mv17@iisertvm.ac.in}
	\affiliation{School of Physics, IISER Thiruvananthapuram, Maruthamala PO, Vithura, Kerala, India 695551}
	
	
	\begin{abstract}
		Constraint satisfiability problems, crucial to several applications, are solved on a quantum computer using Grover's search algorithm, leading to a quadratic improvement over the classical case. The solutions are obtained with high probability for several cases and are illustrated for the cases involving two variables for both  3- and 4-bit numbers. Methods are defined for inequality comparisons, and these are combined according to the form of the satisfiability formula, to form the oracle for the algorithm. The circuit is constructed using IBM Qiskit and is verified on an IBM simulator. It is further executed on one of the Noisy Intermediate-Scale Quantum (NISQ) processors from IBM on the cloud. Noise levels in the processor at present are found to be too high for successful execution. Running the algorithm on the simulator with a custom noise model lets us identify the noise threshold for successful execution.
	\end{abstract}
	
	
	\maketitle
	
\section{Introduction} \label{intro}
    
The satisfiability problem involving conjunctive inequalities is one of the widely encountered problems in database systems, forming a central part of several database problems. Quite a few application areas such as program analysis, scheduling, planning, testing, and verification rely on constraint-satisfaction problems~\cite{SMT}. The Boolean Satisfiability problem (SAT or B-SAT) plays a vital role in such applications. Solvers for such formulations, called the satisfiability modulo theories (SMT-solvers), have thus been an active area of research in the classical theory of computation \cite{SMT1,SMT2,SMT3,SMT4,SMT5,SMT6}. When quantum computers also develop to a stage where it can tackle complex real-world problems and applications, it is likely that solvers for constraint-satisfaction problems, implemented on quantum computers, would be important as well. In this article, we show how some constraint-satisfaction problems can be solved on small-scale quantum information processors. We consider conjunctive formulae of arithmetic inequalities of the form $(X$ \textbf{op} $C)$ and $(X$ \textbf{op} $Y)$, where $C$ is a constant in the domain of $X$; $X$ and $Y$ are attributes/variables from the integer domain, and \textbf{op}  $\in \{<, \leq, =, \neq, \geq, > \}$~\cite{guo}. A conjunctive normal form (CNF) is a product of sums or an AND of ORs. Such a formula is satisfiable by an assignment of variables if it evaluates to true under the particular assignment. 

The general satisfiability problem of checking if a formula evaluates to true under any of the assignments of variables from its domain has been shown to be NP-hard in the integer case~\cite{rosenkrantzhunt}. However, in the integer domain involving $\textbf{OP}_{\neg \neq}$, the satisfiability problem has been shown to be solvable in $O(|S|)$ time where $S$ is the conjunctive inequality to be solved and $|S|$ is its size, i.e., the number of inequalities in $S$. Here, $\textbf{OP}_{\neg \neq}$ is the group of operators excluding the $\neq$ operator, i.e. $\textbf{OP}_{\neg \neq} \equiv \{<, \leq, =, \geq, > \}$~\cite{guo}. The proposed solution in~\cite{guo} employs an approach using directed graphs to express the problem and determining if $S$ is satisfiable or not. In this article, the aim is to find the solutions in addition to checking for satisfiability, i.e., searching for assignments of variables satisfying $S$. This involves searching through all the possible variable assignments and checking for satisfiability. Hence, the problem translates into a database search problem. 

\section{Classical and Quantum approaches}

One classical approach involves checking through all the possible input combinations to see if the inequality is satisfiable by any of them. So, if there are $N$ combinations to be checked for a case with a single solution, then this classical approach involves checking each combination until it is found. In the worst case, all the $N$ cases have to be tried where the $N^{\rm th}$ element is the solution. Hence, on an average, this approach requires $N/2$ operations to find the solution, and the classical algorithm is of $O(N)$~\cite{qkit,qcqi}. While specialized approaches based on heuristics can give improvements for specific search problems with additional structure, the number of steps required for finding the solutions of generic search problems using classical approaches scales as $O(N)$ or faster~\cite{Papadimitriou15881}.

In solving several problems, quantum computers have been shown to be much more efficient than the classical ones. One of the first examples showing this advantage was in the early '90s, which is the Deutsch-Jozsa algorithm \cite{deutschjozsa} which determines whether a function is constant or balanced with just one evaluation of the function. This corresponds to an exponential speed-up over the best known classical algorithm. However, this algorithm solves a problem of not much practical interest. The Shor's algorithm \cite{Shor} in early 1994 showed that quantum computers can efficiently solve the well-known problem of integer factorization with an exponential speed-up over the classical case. Both algorithms, Deutsch - Jozsa as well as Shor's, are based on the quantum versions of the Fourier transform \cite{qcqi}. Quantum simulations, wherein a quantum system is simulated using a quantum computer itself, as proposed by Feynman~\cite{Feynman,qcqi} is another family of problems for which quantum computers can furnish an exponential speed-up over classical ones.

Another class of problems for which quantum algorithms can  provide an advantage over classical ones is database search. Although the quantum database search algorithm, namely Grover's algorithm~\cite{Grover1,Grover2}, provides only a quadratic speedup over classical ones, it does find use in much more problems of practical interest and has a wider range of application \cite{qcqi}. We briefly describe the algorithm below and show how it can be applied to the satisfiability problems we are interested in solving.

\subsection{Grover's algorithm}

The quantum parallelism feature enables a quantum computer to simultaneously evaluate a function on all possible inputs, thereby contributing to the speed up over classical algorithms. This feature is used in the Grover's algorithm to give a quadratic speed-up over the classical case in database search. For the case of a search problem over ${\cal N} = 2^{\tilde{n}}$ alternatives, the quantum processor represented by an $\tilde{n}$ qubit register is initialized in the uniform superposition state 
\[ \frac{1}{\sqrt{\cal N}} \sum_{i=1}^{\cal N} |x_i\rangle = |\psi\rangle,\] 
by performing a Hadamard transform on each qubit. This can be done using $O(\log_{2}{\cal N})$ one-qubit gates \cite{Grover2}. Now, the Grover iteration is performed on this register. 

A Grover iteration consists of two steps - the oracle and the diffusion operator \cite{qkit}. The oracle marks those states of the quantum processor corresponding to combinations of variable values that satisfy $S$ with a negative phase by rotating the phase by $\pi$ radians. Its action can be expressed as $|x\rangle \longrightarrow (-1)^{f(x)}|x\rangle$. The oracle is constructed so that it can \textit{recognize} the solution(s) to the search problem without actually \textit{knowing} them. Hence, $f(x)$ is equal to 1 for the solution states and 0 otherwise. The action of the diffusion operator can be written as $H^{{\otimes}\tilde{n}}(2|0^{\tilde{n}} \rangle \langle 0^{\tilde{n}}| - {\mathbb{I}}_{\tilde{n}}) H^{{\otimes}\tilde{n}} $, which is the same as $(2|\psi \rangle \langle \psi| - {\mathbb{I}}_{\tilde{n}} )$, where $|\psi\rangle$ is the uniform superposition state. It can be shown that this is essentially the \textit{inversion about the average} operation \cite{Grover1,qcqi}. Combining the oracle operation ${\mathcal O}$ and the diffusion operation, the Grover iteration is $G = (2|\psi \rangle \langle \psi| - {\mathbb{I}}_{\tilde{n}} ){\mathcal O}$. Geometrically $G$ can be understood as a rotation in the two-dimensional space spanned by the state $|\phi \rangle$, which is the uniform superposition of the solutions to the search problem and the state $|\phi_\perp\rangle$ which is perpendicular to $|\phi\rangle$. For a search problem with one solution and ${\cal N}$ values to be searched, each Grover iteration increases the amplitude in the desired state by $O(1/\sqrt{\cal N})$. In other words, the state of the $\tilde{n}$-qubit register is rotated and brought progressively closer to the subspace containing the solutions of the search problem. After each iteration the projection of the $\tilde{n}$-qubit state on the target state $|\phi\rangle$ increases by a factor of $1/\sqrt{\cal N}$ as explained in more detail in Sec.~\ref{groveriter}. Thus, the amplitude and hence the probability weight of the desired state reaches $O(1)$ after $O(\sqrt{\cal N})$ repetitions of the Grover iteration \cite{Grover2}. In general, the number of Grover iterations (oracle+diffusion) required for obtaining solutions with high probability is approximately $(\pi/4)\sqrt{{\cal N}/M}$, where ${\cal N}$ is the number of items in the search problem and $M$ is the number of solutions~\cite{qcqi}. 

    \onecolumngrid
    \begin{center}
    \begin{figure}[h]
    \includegraphics[width=0.99\textwidth]{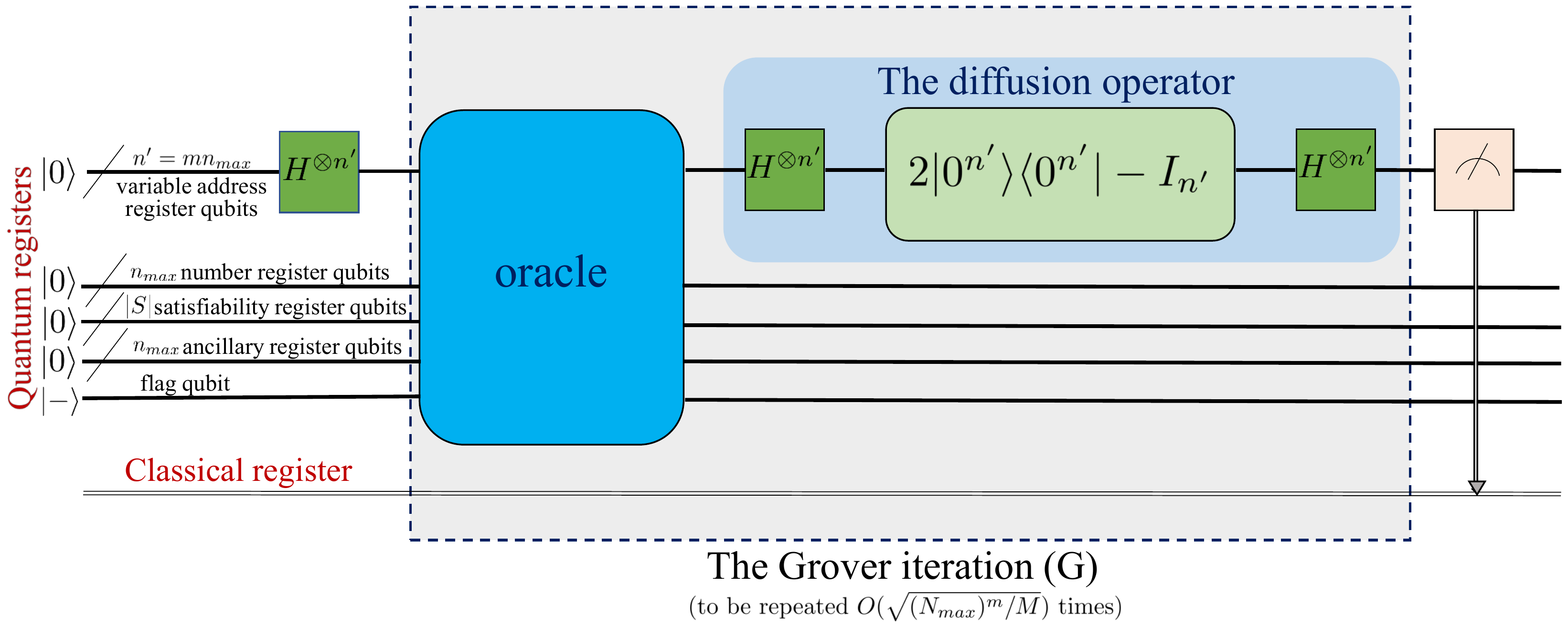}
    \caption{The Grover's algorithm circuit.\label{Grover_fig}}
    \end{figure}
    \end{center}
    \newpage
    \twocolumngrid
    
In the case of the satisfiability problem we are considering, the variables are first put in a superposition of all the values they can be in. For example, let the conjunctive formula to be checked for satisfiability be 
\begin{equation}
    \label{example1}
    S = \{(X < 8) \wedge (Y = 4) \wedge (X > Y)\}.
\end{equation} 
This is a conjunction of 3 inequalities where the first and the second one involve $X$ and $Y$ alone respectively, while the third one involves both. Initially, if we consider $X$ and $Y$ to be bit strings of the same length, $n$, we have $N = 2^n$ values for each of these variables. These variables are first put in a uniform superposition of all these values that they can take by performing a Hadamard transform on each qubit. We need $n$ qubits for each variable register to create this equal superposition of $N$ states. If we consider the domain of $X$ and $Y$ as all the integers from 0 to 15 (both included), we will need 4 qubits in each register. So, in general, for $m$ variables, we require $mn_{\max}$ qubits to create all the variable address registers, where $n_{\max} = \log_{2}N_{\max}$, with $(N_{\max}-1)$ being the largest integer value that the variables can take. All the other integers are also expressed in the same binary string pattern as $(N_{\max}-1)$, by filling in zeroes at the empty places. Hence, the number of values to be searched is $(N_{\max})^{m}$, which can be solved in $O(\sqrt{(N_{\max})^{m}/M})$ iterations, where $M$ is the number of solutions. The Grover's algorithm implementation for the general case of $M$ solutions is summarized in Fig. \ref{Grover_fig}.

In the present example, it is easy to see that the values of $X$ and $Y$ for which the given inequality is satisfied are (4, 5), (4, 6) and (4, 7) [(0100, 0101), (0100, 0110) and (0100, 0111) in binary  format], expressed in the form $(Y, X)$. The total number of values to be searched is $16^2 = 256$ and out of these, we have three solutions. The states corresponding to these combinations of variable values will be flipped initially by the oracle. The diffusion or the amplitude amplification operator then raises the amplitude of these states (while diminishing that of the others) by inverting about the mean, making their probabilities higher so that a measurement yields one of these states preferentially (or with greater probability), as required.

\section{Implementation}

Considering an example with two variables, the implementation will consist of a total of five quantum registers. First, we create the address register to store the variable values. In the example we are considering, both $X$ and $Y$ are taken to be four-bit numbers and so we require $2 \times 4 = 8$ qubits in total for the two-variable address register. Initially, all the qubits are in the $|0\rangle$ state. Next, we have the satisfiability register qubits to store the information regarding each of the inequalities in the conjunctive formula for satisfiability. The number of satisfiability register qubits required is equal to the number of inequalities in the formula, which is 3 in this case. To store the values of the constants (assumed to take values in the domain of the variables themselves), a number register of the same size as the variable registers would be required. Finally, we need a single flag qubit, whose phase can be flipped when all the inequalities are satisfied (i.e., the case when all the satisfiability qubits become $|1\rangle$). In this way, the solution states get marked in the oracle operation. The total number of qubits required for implementing the example we have considered is equal to 4 (for the X register)+4 (for the Y register)+4 (for the number register)+3 (for the satisfiability register)+1 (flag qubit) = 16. As we see in the following, the implementation would require additionally $n=4$ ancilla qubits. Adding the 4 ancilla qubits we require a total of 20 qubits for this particular example. 

In case $S$ is not satisfiable by any of the input combinations, none of the states gets marked, and we get an almost uniform probability distribution. A similar probability distribution occurs in the case of a tautology too, as all states get marked by the Grover oracle. These extreme cases can be distinguished by making appropriate changes in the implementation so as to finally make a measurement on the answer qubit (instead of on the variable registers), which will be in the $|1\rangle$ state for all values in the case of a tautology and the $|0\rangle$ state if not satisfiable. 

\subsection{Elementary operations in the oracle}
    
The oracle is a major component of the Grover's algorithm. It picks out the solution(s) from all the input combinations. Here, the oracle operation consists of recognizing the $X$ and $Y$ values which satisfy all the inequalities in $S$ simultaneously. In this implementation, methods (functions) are defined for checking for the `less than' ($<$) and `equal to' ($=$) comparisons. The `greater than' ($>$) comparison can be performed by reversing the inequality and using the function for $<$ itself. The $\leq$ and $\geq$ relations can also be realized by simultaneously applying two of these comparisons. The comparison modules have been elaborated here for the case of positive integers. Comparisons between positive and negative integers can be done in the same way by including one more qubit each for every variable that carries the information regarding the sign. Alternatively, the comparisons can be done in the positive domain alone by providing a constant offset to the whole problem (and every variable and constant involved) by the smallest negative value in the particular case. After the comparison, the variables and constants can finally be shifted back to the actual range so that the solutions correspond to the actual problem. In other words, a problem in the domain $(-n_1, n_2)$ can be reformulated as one in $(0, n_1+n_2)$, without any loss of generality. In this case, extra qubits will not be required to encode the information regarding the signs of the literals. 

The oracle itself is constructed by combining together the elementary functions like $<$ and $=$. First, we will discuss the implementation of the elementary functions in detail before proceeding to see how they can be combined to form the oracle. 

\subsubsection{The $<$ comparison:}

    \begin{figure}[ht!]
    \includegraphics[width=0.99\linewidth]{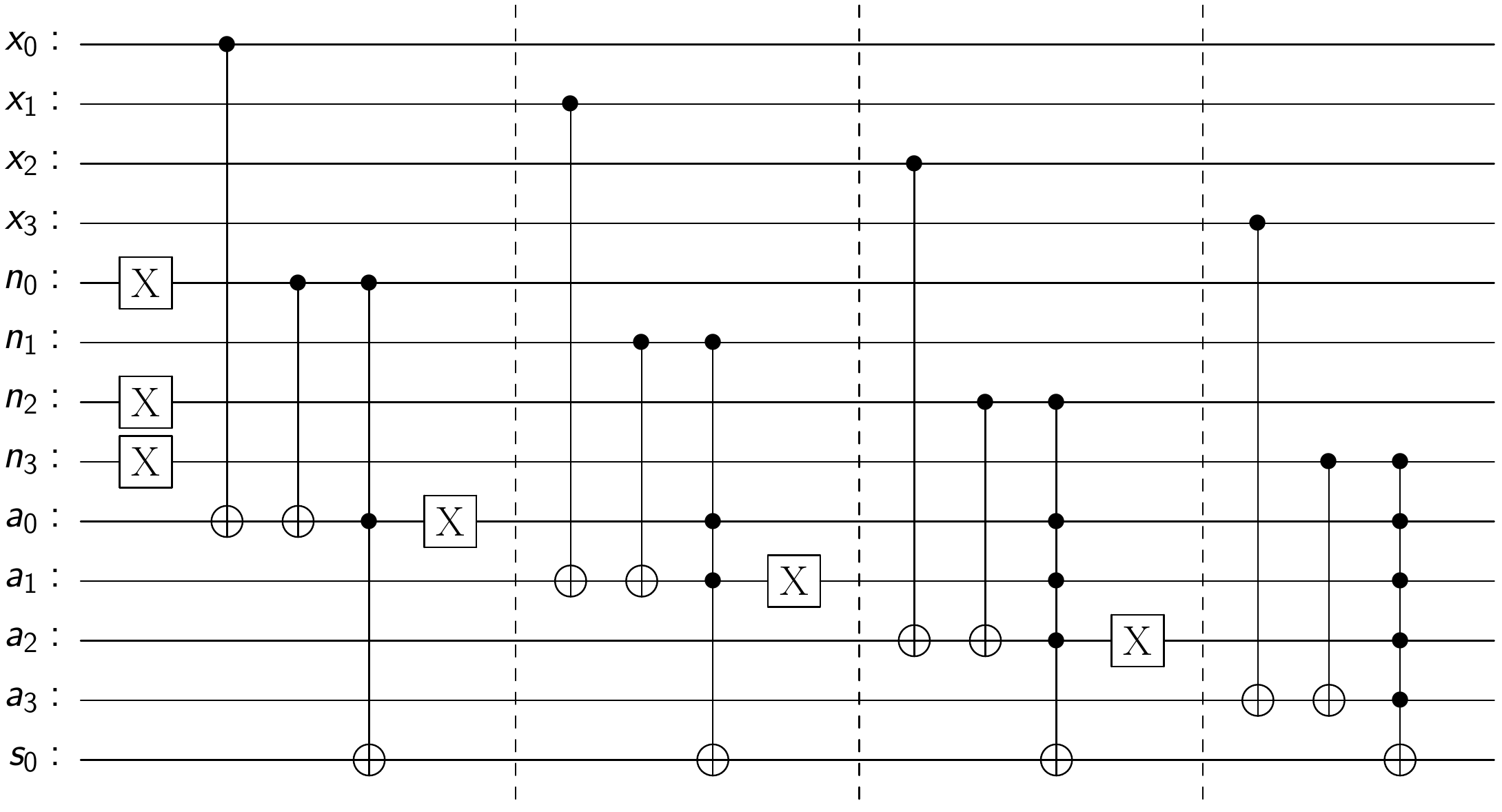}
    \caption{The circuit for $<$ comparison for 4-bit numbers. The number being compared with the constant in the number register (the number 11 here) needs to be encoded onto the $X$ register. Here, each $x_i$($y_i$), $n_i$ and $a_i$ denote the indices of the qubits in the variable, constant and ancilla qubit registers, respectively. The label $s_0$ denotes the  qubit from the satisfiability register that is used to store the information about the satisfiability of this component circuit. The barriers shown in the figure are only to improve the readability. They are removed before the actual implementation since optimization does not work across the barriers. \label{lessthan_fig}}
    \end{figure}
    
The circuit used for the $<$ comparison is shown in Fig. \ref{lessthan_fig}. This is one way of comparing if a variable is less than another or a constant.  The IBM Qiskit Circuit Library has an integer comparator operator (qiskit.circuit.library.IntegerComparator) defined~\cite{c_lib}. A quantum bit string comparator (QBSC) is also discussed in~\cite{QBSC}, which can provide information regarding whether the integers are equal, or if not, which one is greater or smaller. An application of this QBSC in the Grover's search algorithm has also been mentioned. However, it may be noted that such multi-purpose comparison operators come with their own implementation overheads and may require additional ancilla qubits. We therefore discuss independent implementations of each of the elementary functions that use a minimal number of ancilla qubits which can be re-used as well. 

 In Fig.~\ref{lessthan_fig}, the comparison is shown to check for the case $(X < 11)$. Here, the constant $11$ is encoded as $1011$ onto the number register ($n_j$). The $X$ address register contains the input superposition (an equal superposition of all possible values ($0$ to $15$) during the first iteration) and the Hadamard gates that put the register in this state are not shown in the figure here. All values of $X$ will be simultaneously checked for the validity of the inequality and the ones satisfying it will be marked. To check for the validity of such a relation between two variables, as in the case of $(X < Y)$ in $S$ from Eq.~\eqref{example1}, the second register used will be the one corresponding to the second variable (the $Y$ register). The comparison uses 4 ancilla qubits as well. Note that the number of ancillary qubits used does not typically figure in the computation of the  complexity of the algorithm as the analysis involves only the qubits in the data registers~\cite{qcqi}. In our implementation, the ancilla register adds as many qubits as required to store one of the input variables. 
 
 The logic used in the circuit in Fig.~\ref{lessthan_fig} is to compare each bit in the binary representation of the values. The comparison starts from the Most Significant Bit (MSB) position. Let us consider the instance of comparing 8 (1000 in binary) and 11 (1011 in binary). An XOR gates checks if two qubits are different or not. The XOR is implemented by a pair of `controlled-X' (CX) gates, with the first of the two qubits being compared as control for the first CX and the second as the control for the second CX. The ancilla qubit corresponding to the pair of qubits being compared (the MSB qubits in this case) is the target. A similar procedure for comparison has been used in an example in Section 3.8 of~\cite{qkit}. The ancilla qubit will be set to $|1\rangle$ only if both the qubits being compared correspond to different bit values. For our example, the MSB bit is 1 for both the numbers being compared. The XOR does not flip the corresponding ancilla and the comparison shifts to the next position. As these two bits are again the same, the bit values in the next position are compared. Since these are different, the circuit checks if the 1 occurs in the case of the constant (the number 11), since the required comparison is $X<11$. Since inequality holds true for $X = 8$ the  corresponding satisfiability register qubit in the oracle implementation (denoted as $s_0$ in the figure) is set to 1. Setting the value of the satisfiability qubit is done by multi-controlled Toffoli gates having the corresponding qubits from the number register and the ancilla qubit register as controls. The ancilla qubit holding the information regarding whether the bits are the same or different is inverted after each comparison (by applying an $X$ gate, as shown in the figure) so that its value does not affect the subsequent comparison(s) that, in turn, are controlled by all the previous ancilla qubits. 
 
 The uncomputation part, which takes all the ancilla qubits back to their initial state (the $|0\rangle$ state) is not shown here, but it is the same set of operations done in reverse order because all the operations used here are their own inverses. The satisfiability register qubits are uncomputed only after the execution of all the elementary operations in the conjunctive formula. The information encoded in these qubits is required to check the simultaneous satisfiability of all the component circuits, which is an operation done only after all component circuits are executed. The ancilla qubits may have gotten entangled with the data qubits during the course of the computation. If residual, garbage values are allowed to remain in the ancilla qubits, decoherence of the ancilla qubits  measurements done on them subsequently can lead to a collapse of the superposition in the data qubit registers also because of their mutual entanglement. Uncomputation not only removes entanglement between the ancilla qubits and the data qubits, it also allows for their reuse~\cite{uncompute}.  

\subsubsection{The $=$ and $\neq$ comparison:}

    \begin{figure}[ht!]
    \includegraphics[width=0.99\linewidth]{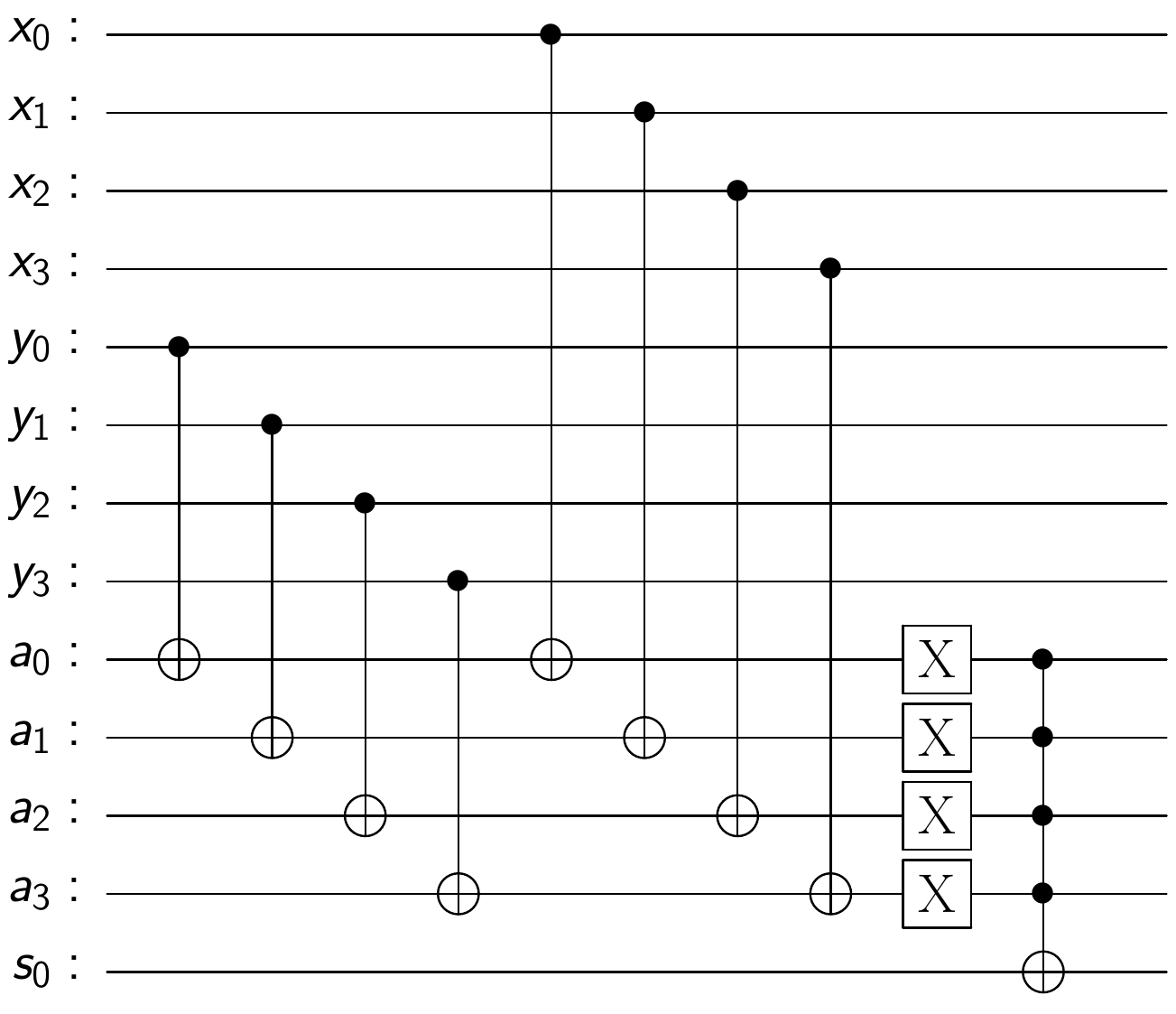}
    \caption{The circuit for $=$ comparison for 4-bit numbers. The numbers  being compared are stored in the $X$ and $Y$ registers. Here, each $x_i$, $y_i$ ($n_i$) and $a_i$ denote the indices of the qubits in the variable, (constant) and ancilla qubit registers, respectively. $s_0$ denotes the  qubit from the satisfiability register that is used to store the information about the satisfiability of this component circuit.\label{equal_to_fig}}
    \end{figure}

Fig.~\ref{equal_to_fig} shows the circuit used for the `equal-to' (and by extension `not-equal-to') comparison. We have shown the comparison between  values in two variable registers (i.e., the comparison $(X = Y)$). The values of $X$ and $Y$ being compared need to be encoded into the respective registers. They can also be in a superposition of all the possible states. The number of ancilla qubits required is equal to the length of a variable address register (4 in this case). As in the $<$ comparison, the corresponding bits in each variable are checked for equality using XOR gates. In the cases where the bits are the same, the ancilla qubit corresponding to that particular position will remain in the $|0\rangle$ state itself. We invert all the ancilla qubits in the last step (so that they are all in the state $|1\rangle$ for the solution case) and then implement a multi-controlled Toffoli gate with all these qubits as controls and the satisfiability qubit as the target. The uncomputation is again the same set of operations done in the reversed order. Note that the circuit for $\neq$ comparison will require an additional $X$ gate on the answer qubit after the equality comparison so that the cases in which the values are not equal will get marked.

Checking if a variable is equal to a constant can be done without using a separate register to hold the value of the constant. For instance, suppose we want to check if $X=4$ (0100 in binary). The $X$-register may be in an arbitrary superposition state. We implement $X$-gates (NOT-gates) on all qubits of the $X$-register where the constant has bit value 0 ($x_0$, $x_1$ and $x_3$ in this case) and then implement a multi-controlled Toffoli gate with all the qubits in the $X$-register as controls and the satisfiability register qubit as the target. Only the state $|0100\rangle$ of the $X$-register is converted to $|1111\rangle$ by the NOT-gates we added, and the multi-controlled Toffoli, in turn, has a non-trivial action on the target only when the control qubits are in the $|1111\rangle$ state, thereby implementing the comparison of the variable with the specified constant. A similar comparison had been shown in \cite{hint_qkit}.

\subsubsection{Toffoli gate}

    \begin{figure}[ht!]
    \includegraphics[width=0.99\linewidth]{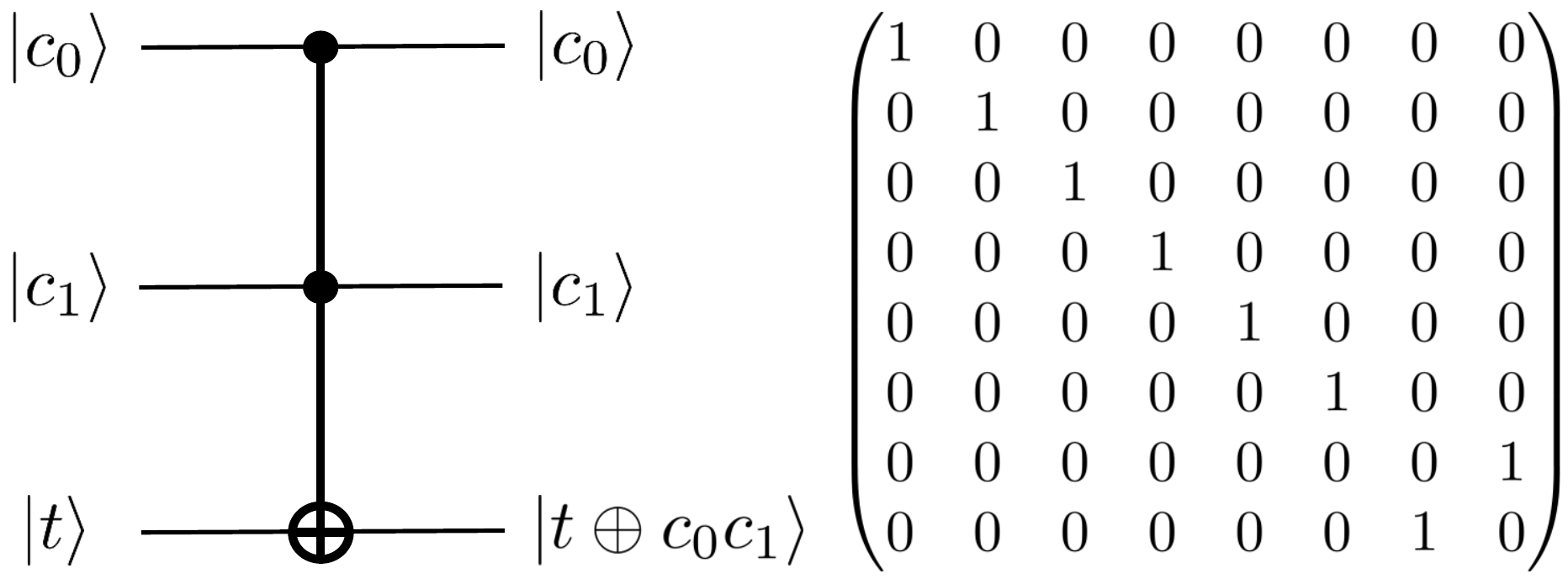}
    \caption{The Toffoli gate circuit and the unitary describing its action.\label{Toffoli_gate}}
    \end{figure}

The Toffoli gate (Fig. \ref{Toffoli_gate}) is a controlled-controlled-NOT gate (or the CCX gate) \cite{qkit}. It is a unitary operation acting on 3 qubits. The first two qubits are the control qubits, and the $3^{\textrm{rd}}$ one is the target qubit which is flipped if both the control qubits are set to $|1\rangle$. It is a reversible gate, being its own inverse~\cite{qcqi}. The Toffoli gate action can be extended to multiple controls and multiple targets. The multi-controlled Toffoli (MCT) gate, performing the C$^{{\otimes}n}$X operation has been used extensively in the circuits described here. In this case, the target qubit is flipped if all the control qubits are set to $|1\rangle$. 

In general, for two $n$-bit numbers, the `$<$' comparison would require $2n$ CX gates and $(n-1)$ X gates in addition to C$^{{\otimes}m}$X gates where $m \in \mathbb{Z}$ and ranges from $2$ to $(n+1)$. Without using any imperfect relative phase Toffoli (RTOF) gates or ancillary qubits, decomposing these MCT gates (C$^{{\otimes}m}$X gates) into elementary gates would require $O(m^2)$ basic operations ~\cite{RTOF_M, no_ancilla1, qcqi}. Similarly, the `=' comparison would require $2n$ CX gates, $n$ X gates and a C$^{{\otimes}n}$X gate. For instance, the oracle for the `$<$' operation shown in Fig.\ref{lessthan_fig} for two 4-bit numbers can be implemented using 160 U3 gates and 156 controlled-X (CX) gates, when unrolled using the U3+CX gateset. (U3 is the generic single qubit rotation gate in Qiskit.) Calculating the cost by setting the cost of the controlled-X gate to be 10 times that of the general unitary gate, the cost for this operation is 1720. Similarly, the `=' comparison (Fig. \ref{equal_to_fig}) will require 44 U3 gates as well as 44 CX gates (cost: 484). However, the cost was calculated for the case of using ideal gates without employing any cost reduction techniques.
The implementation of the C$^{{\otimes}n}$X operation using the limited gate set of real NISQ \cite{NISQ} devices has been discussed in Sec.~\ref{optimization}. For instance, the optimization strategies for various MCT gates using ancilla and relative phase gates have been elaborated in \cite{RTOF_iX}, employing roughly $O(n)$ basic operations.

\subsection{Grover Iteration}
\label{groveriter}

The first step of the Grover iteration, namely the oracle, is constructed by combining the elementary operations in the formula $S$ in series. The registers holding the variables, constants and the ancilla qubits are common to all elementary operation circuits. Component circuits do not alter the quantum states of the registers holding the variables and constants since the corresponding qubits are only used as controls in the implementation of the respective operations. This means that these registers can be used as it is in all the component circuits, with appropriate re-initialization of the qubits storing the constants. Ancilla qubits, on the other hand, are changed by each component circuit, but the uncomputation that is built into each such operation allows for their re-use. The satisfiability qubits of each component circuit cannot be re-used since it contains the information about which components of the input superposition state of the data registers satisfy the particular operation that is part of $S$. 

In those cases where components of the superposition in the variable registers satisfy an elementary constituent inequality of $S$, the satisfiability qubit of that circuit is set to the state $|1\rangle$. Whenever the satisfiability qubits of all the elementary circuits  that constitute the conjunctive formula are set to $|1\rangle$ it means that the formula is satisfied. A flag qubit is now added which is initialized in the state $|-\rangle = (|0\rangle - |1\rangle)/\sqrt{2}$. The quantum state of the registers at this point can be written as
\begin{eqnarray} 
    \label{state1}
    |\psi'\rangle & = &  \!\!\!\! \sum_{x_i,y_i=0}^{1} |x_1 \ldots x_d\rangle   |y_1 \ldots y_d\rangle  |n_1 \ldots n_d\rangle  \nonumber \\
    && \quad \otimes|{\mathrm{ancilla}}\rangle |s_1^{x_i,y_i}\rangle  |s_2^{x_i,y_i}\rangle  |s_3^{x_i,y_i}\rangle  |-\rangle_{\mathrm{f}},
\end{eqnarray}
where $|x_1 \ldots x_d\rangle$ and $|y_1 \ldots y_d\rangle$ denote the variable registers which are in a superposition of all possible values and $|n_1 \ldots n_d\rangle$ denotes the register holding the values of the constants. The ancilla register is also shown explicitly in the state given above even if at the end of the computation the state of this register is restored to its initial one and is therefore irrelevant. The satisfiability qubits from each component circuit, assuming that there are three such circuits, are shown as $|s_j\rangle$ with their dependence on the values $x_j$ and $y_j$ of the input register shown explicitly in superscript. The last qubit is the flag qubit initially in the state $|-\rangle$. 

The final step in the oracle action is the application of a multi-controlled Toffoli gate with all the satisfiability qubits $|s_j\rangle$ as controls and the flag qubit as the target. When all $s_j=1$, the Toffoli gate acts as 
\[ |-\rangle_{\mathrm{f}} \rightarrow -|-\rangle_{\mathrm{f}} = (-1)^{s_1s_2s_3}|-\rangle_{\mathrm{f}}, \]
where $s_1s_2s_3$ denotes the product of the bit values corresponding to the satisfiability qubit states. This converts the state in Eq.~\eqref{state1} to
\begin{eqnarray} 
    \label{state2}
    |\psi''\rangle & = & \!\!\!\! \sum_{x_i,y_i=0}^{1} (-1)^{s_1s_2s_3}|x_1 \ldots x_d\rangle   |y_1 \ldots y_d\rangle  |n_1 \ldots n_d\rangle  \nonumber \\
    && \quad \otimes|{\mathrm{ancilla}}\rangle |s_1^{x_i,y_i}\rangle  |s_2^{x_i,y_i}\rangle  |s_3^{x_i,y_i}\rangle  |-\rangle_{\mathrm{f}}.
\end{eqnarray}
In writing the state above we have explicitly shown the phase kickback produced by the final Toffoli gate that marks the solution states with a phase of -1 thereby completing the oracle action. Alternatively, the flag qubit can be set to the $|1\rangle$ state initially, instead of the $|-\rangle$ state; and the multi-controlled Z (MCZ) gate can be used instead of the MCT gate to check for the simultaneous satisfiability of all the inequalities. This also has the similar phase kickback effect on the solution states.

Since the register holding the constants and the ancilla register qubits were re-used at each stage by uncomputing them, the total number of qubits required will not scale with the number of inequalities in $S$ (provided the number of variables remain the same). This has the advantage of reducing the number of qubits used, but the qubits belonging to the number register and the ancillary register must both be uncomputed after each operation (since they are used as a common resource for all the operations). Reusing these registers means that computation of each of the qubits in the satisfiability register cannot be done in parallel and there will also be a small cost involved in re-initializing the number register at each step for each constraint. After all these operations and the phase kickback operation, the satisfiability register qubits alone remains to be uncomputed. The operations (used for checking for the validity of the inequalities) which were performed earlier can be repeated to uncompute the satisfiability register qubits and bring them back to their initial state (the $|0\rangle$ state). After this uncomputation of the satisfiability register qubits, the state in Eq.~\eqref{state2} becomes
\begin{eqnarray} 
    \label{state3}
    |\psi'''\rangle & = & \!\!\!\! \sum_{x_i,y_i=0}^{1} (-1)^{s_1s_2s_3}|x_1 \ldots x_d\rangle   |y_1 \ldots y_d\rangle  |n_1 \ldots n_d\rangle  \nonumber \\
    && \quad \!\!\!\!\! \otimes|{\mathrm{ancilla}}\rangle |{\mathrm{uncomputed\:satisfiability}}\rangle  |-\rangle_{\mathrm{f}},
\end{eqnarray}
where $|{\mathrm{uncomputed\:satisfiability}}\rangle$ is the state of the satisfiability register qubits after uncomputation.

In all the operations here, the variables and constants were expressed as binary strings having the same length by filling in zeroes in empty places for numbers with a smaller length since the comparison circuits we use call for registers of equal length. 

    \begin{figure}[ht!]
    \includegraphics[width=0.99\linewidth]{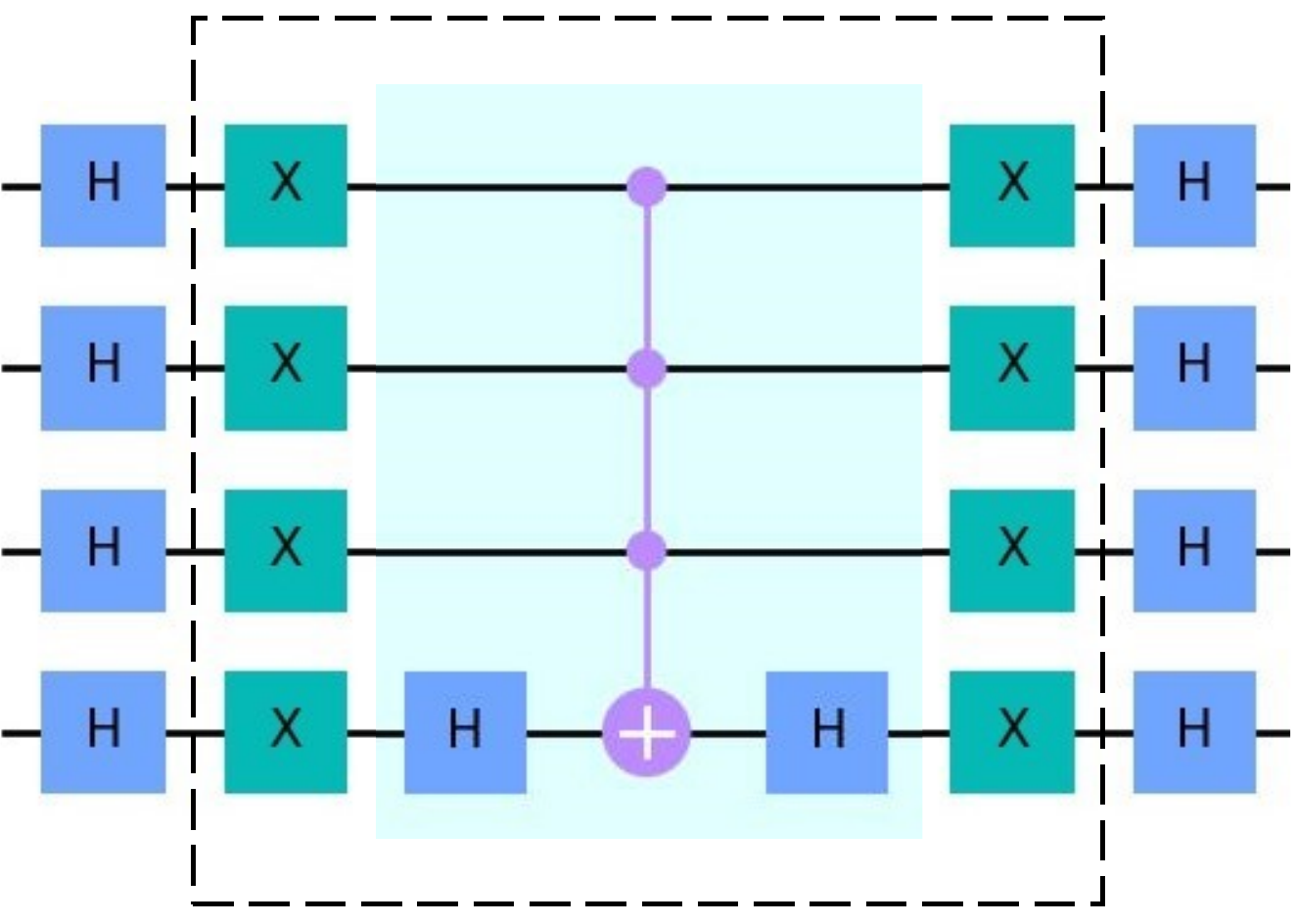}
    \caption{The implementation of the diffusion operation on a 4-qubit register.\label{diffusn_fig}}
    \end{figure}
    
The second step of the Grover iteration is the diffusion (or the amplitude amplification) operator that is applied on the variable registers. This operation, $ H^{{\otimes}n}(2|0^{n} \rangle \langle 0^{n}| - {\mathbb I}_{n})H^{{\otimes}n}$, can be implemented as shown in Fig.~\ref{diffusn_fig} for a 4-qubit case~\cite{qcqi}. In general, the number, $n$, of qubits on which the diffusion operator acts is the total number of qubits in all the variable registers put together. Here, the first set of Hadamard gates would have converted the initial uniform superposition state $|\psi\rangle$ of the variable registers to one in which all qubits are in the  $|0\rangle$ state. The oracle has however marked certain components of the superposition (the solutions) with a phase of -1 in the state $|\psi'''\rangle$ from Eq.~\eqref{state3}. This difference between $|\psi\rangle$ and $|\psi'''\rangle$ is amplified by the diffusion~\cite{qcqi}. The conditional phase shift operation $(2|0^{n} \rangle \langle 0^{n}| - {\mathbb I}_{n})$ is carried out by the part of the circuit in the dotted box \cite{qcqi}. The gate operations in the shaded part comprise the multi-controlled Z (MCZ) operation, where the identity $HXH = Z$ has been used. After sufficient number of iterations,  measurement of the variable register leads to one of the solution states with high probability. Each such solution is stored in a classical register. By repeating this whole process, all the solutions can be identified provided the number of solutions is small compared to the total number of possible inputs $(M\ll N)$. 

\begin{figure}[!ht]
    \includegraphics[width=0.99\linewidth]{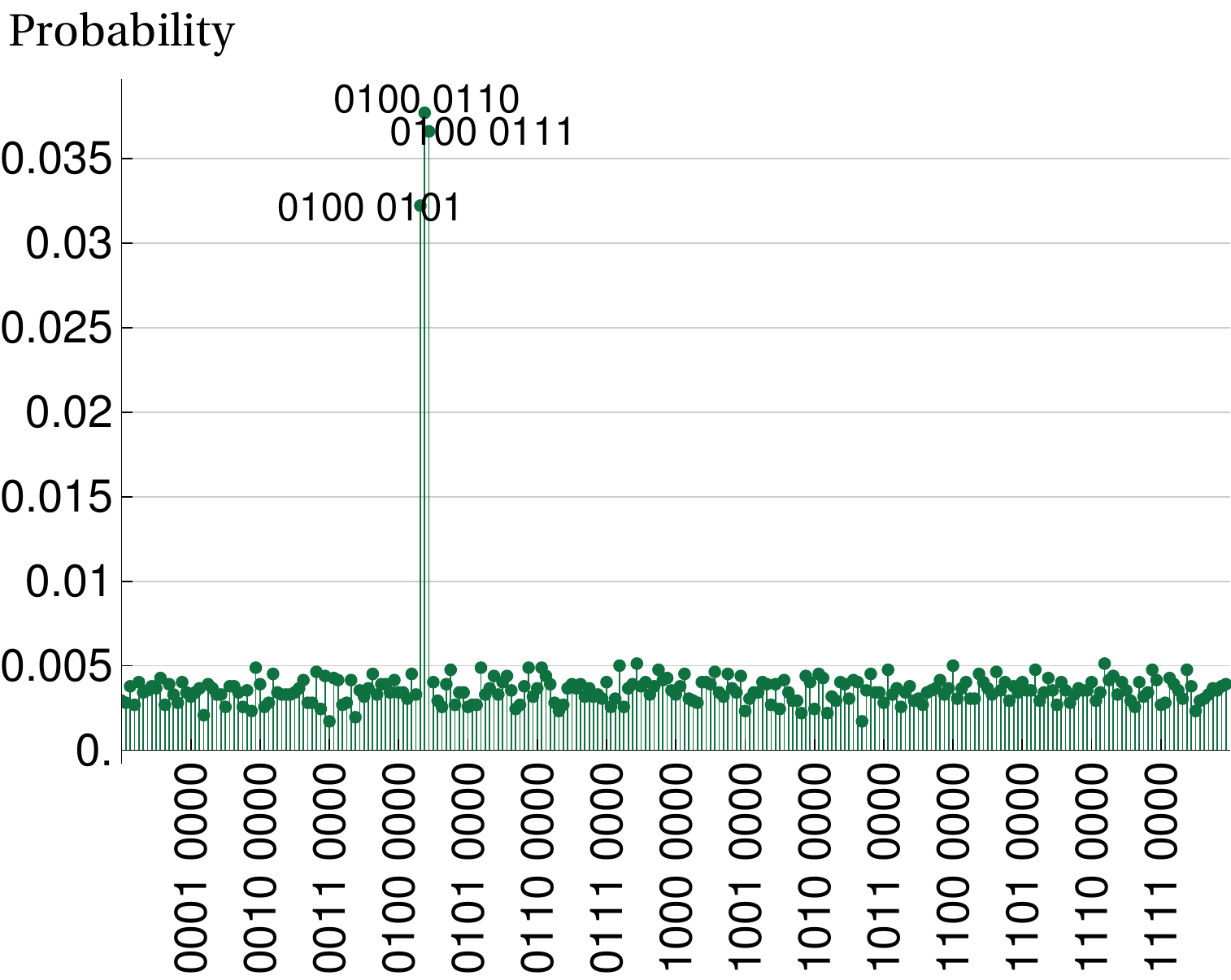}
    \caption{The final probability distribution for the case $S = \{(X < 8) \wedge (Y = 4) \wedge (X > Y)\}$, giving high probabilities for the 3 solutions. The execution was performed on the \textbf{ibmq\_qasm\_simulator} and was run 8192 times to get the probability distribution. $P_{\mbox{tot}} = 0.107.$ \label{prob_dist_fig}}
    \end{figure}

Returning to the specific example we were considering with $N = 256$ and $M = 3$, the expected number of iterations for sufficiently amplifying the amplitude of the solutions is approximately 8. However, the solutions were obtained with considerably high probability (total probability of finding a solution, $P_{\mbox{tot}} = 0.107$) in just one iteration itself. Although only a single iteration was performed, the circuit execution was repeated 8192 times, making the solution states easily distinguishable in the final probability distribution. Increasing the number of iterations leads to a circuit of greater depth and greater cost as well, which is not desirable. Still, without incorporating this high cost circuit, it is possible to easily obtain the solutions in a single iteration alone, by increasing the number of execution repetition shots. Fig.~\ref{prob_dist_fig} shows the final probability distribution after one of the implementations of the circuit on the \textbf{ibmq\_qasm\_simulator} (provided by the IBM Quantum Experience \cite{IBMQ}), which supports up to 32 qubits. The coding had been done in Python using QisKIT v0.26.0 - IBM's quantum computing SDK, using IBM Quantum Experience cloud-based access.

\section{Circuit cost reduction and optimization} \label{optimization}

The circuit we have designed for the conjunctive formula in Eq.~\eqref{example1} works very well on the IBM simulator giving a sharp increase in the probability for measuring the solution states at the output as seen in Fig.~\ref{prob_dist_fig} after only one Grover iteration. However, the circuit still tests the limits of computations that can be implemented in the NISQ processors available today. In order to successfully implement, we need to consider various circuit optimization and error mitigation steps. A survey on synthesis and optimization of reversible circuits has been done in~\cite{survey}, wherein several algorithmic techniques have been reviewed. It also discusses the need to devise and use methods suitable for a particular device (technology mapping), by taking into account its specific limitations, since the physical implementation on a device is dependent on its Hamiltonian. Several benchmarking and software tools have also been elaborated. Noise-adaptive compiler mappings for NISQ computers, based on several comparative studies, is discussed in~\cite{noise_adaptive}. 

One approach to reducing circuit cost is to use alternative gates, sometimes at the cost of introducing more ancilla qubits. The circuit cost can be brought down considerably by replacing the multi-controlled Toffoli (MCT) gates we have used with alternate low-cost options. Variants of the Toffoli gates that involve imperfect phases may be used instead of the usual Toffoli gates in cases where they are uncomputed, so that these phases do not affect the circuit implementation~\cite{NISQ_Satoh,c_lib,RTOF_1,RTOF_2,RTOF_iX}. These gates fall under the general class of relative phase Toffoli gates (RTOF). The RTOF$_M$ gate by Margolus \cite{RTOF_M} has been proven to be optimal in~\cite{optimal}. This Margolus gate, available in the Qiskit Circuit Library~\cite{c_lib} as the RCCX gate (a relative phase alternative to the controlled-controlled X gate), can be used as an imperfect Toffoli gate. Another realization of an imperfect Toffoli gate is by sandwiching a CZ gate (between the second control and the target) between controlled-Hadamard (from the first control to the target) gates~\cite{qkit,CZCH}. Both these alternative realizations reduce the cost of the Toffoli gate by half, as can be checked by unrolling them into the general unitary gate and the controlled-X gate basis. The Qiskit Circuit Library also includes the C3XGate (for C$^{{\otimes}3}$X operation) and the C4XGate (for C$^{{\otimes}4}$X) methods for performing similar operations, with the C4XGate based on an implementation given in~\cite{RTOF_M}. The RCCCX gate, based on the implementation discussed in~\cite{RTOF_iX}, belongs to the RC3XGate class. 

For performing general C$^{{\otimes}n}$X gates, the Qiskit library has the MCXVChain gate, which uses a V-chain of CX gates. However, this requires additional qubits as ancilla. Alternative methods on realizing Toffoli gates with no ancilla has been discussed in \cite{no_ancilla1} and \cite{no_ancilla2}. Although these have been shown to be beneficial, implementing these gates on actual quantum processors may be both difficult and imperfect, as already mentioned in \cite{no_ancilla1} itself. This is due to the complications involved in decomposing the rotation gates with small angles used in this implementation into fault-tolerant gates. Linear phase-depth (where phase depth is defined as the minimum number of cycles required to execute all the Z$_N$ gates) ancilla-free decomposition methodology for the MCT gate using the (Clifford+Z$_N$) quantum gate library (Z$_N$ being the $N^{\mathrm{th}}$ root of the Pauli's Z-gate, whose action is to flip the phase of the qubit by an angle of $\pi/N$) has been proposed in~\cite{opt_Zn}. However, for achieving low phase-depth, this proposal also has to use ancilla qubits. The blog post by Craig Gidney~\cite{mct_blog} also briefly discusses some methods of constructing such large CNOT gates. Using ancilla or workspace qubits is one of the more efficient ways of reducing circuit cost effectively. However, with the limitations on the number of qubits available in NISQ processors, further advances and improvements in the hardware are required before these strategies can be employed.

\begin{table*}[t]
\centering
 \scalebox{1.2}{
\begin{tabular}{ | >{\centering\arraybackslash}m{5em} | m{3em} | m{3em} | m{3cm} | m{3cm} | m{3cm} |} 
 \hline
 Bitstring length of the variables & No. of clauses & No. of qubits & \makecell{No. of U3 gates \\ ($N_{U3}$)} & \makecell{No. of CX gates \\ ($N_{CX}$)} & \makecell{Cost \\ ($N_{U3} + 10 \times N_{CX}$)} \\
 \hline
 3-bit & \makecell{2 \\ 3} & \makecell{15 \\ 16} & \makecell{85 - 193 \\ 212 - 272} & \makecell{54 - 166 \\ 182 - 248} & \makecell{625 - 1853 \\ 2032 - 2752} \\
 \hline
 4-bit & \makecell{2 \\ 3} & \makecell{19 \\ 20} & \makecell{162 - 565 \\ 607 - 810} & \makecell{114 - 498 \\ 536 - 722} & \makecell{1302 - 5545 \\ 5967 - 8030} \\
 \hline
\end{tabular}}
\caption{The qubit and gate count for conjunctive formulae involving 2 variables for different number of clauses and bit strings of different lengths are listed in the table. The values of circuit cost were obtained for randomly chosen conjunctive formulae ($S$) and these values must be taken as  rough estimates. Gate cancellation and further optimizations depend on the exact gates used, which, in turn, depend on the exact form of $S$. The cost for inequalities of the form $(X$ \textbf{op} $C)$ will differ from those of the form $(X$ \textbf{op} $Y)$. Since the cost for the `$=$' operation is lower than that for the `$<$' or `$>$' operations, the minimum cost is obtained for cases with the maximum number of `$=$' clauses and the maximum cost for the cases with maximum `$<$' or `$>$' clauses in $S$, preserving the validity of $S$.}
\label{tab1}
\end{table*}

We chose to unroll the circuit into the basis composed of the general unitary gate ($U(\theta, \phi, \lambda)$) and the controlled-X gate for the cost calculation. The optimization level of the Qiskit transpiler was set to the maximum level of 3. The circuit transformations (or equivalently, transpiler passes) varies with each level of optimization with each level  containing all the previous level optimizations also. At each optimization level, the passes vary, due to the corresponding, pre-defined pass managers. Level 0 is a special case compatibility only pass and does not involve any explicit optimizations and it just maps the circuit to the back-end. It is used mainly in characterization experiments, where it is desirable to have no optimizations applied by the transpiler. Level 1 (the default level) is for light-weight optimization that collapses adjacent gates wherever possible. Level 2 is a medium-weight optimization in which exploiting gate commutation relationships, cancellation of as many gates as possible is done. It also generates a layout of gates that is optimized for the noise in each device. Level 3 is for heavy-weight optimization, that includes the block collection and optimal re-synthesis pass and redundant gate removal passes as well. Generally, higher fidelity and lower depth is achieved with higher levels of optimizations, but at the expense of longer transpilation time \cite{Qkit_Yt, Qkit_doc}. The cost of the circuit was roughly calculated by setting the cost of the controlled-X gate to be 10 times that of the general unitary gate. Several cost reduction techniques had been discussed in some of the IBM Quantum Challenge Fall 2020 participants' write-ups ~\cite{adrien_soln, beit_soln}. In the cost reduction we used, all the MCT gates except the one checking for the satisfiability of all the clauses were replaced with imperfect ones, since these were uncomputed later on. The MCT gates with 2 and 3 controls were replaced with the RCCX and RC3X gates. The flag qubit was used as an ancilla for the decomposition of the C$^{{\otimes}4}$X and C$^{{\otimes}5}$X gates. The MCT gate used in the diffusion was also replaced with a low-cost variant composed of RCCX and RC3X gates and a single Toffoli gate, using the ancillary qubits for the decomposition. This kind of decomposition was found to be more cost-efficient than the V-Chain of CX gates which uses more ancilla. The inverses of these imperfect gates were used in the uncomputation of all these operations. With these optimizations done, the circuit cost reduced significantly, by roughly 53-56\% for the specific case of 3-bit numbers. For the case of $S = \{(X < 5) \wedge (Y = 6)\}$, the cost came down from 2889 (289 U3 gates + 260 CX gates) to 1351 (151 U3 gates + 120 CX gates), which is a reduction by around 53\%. Similarly, for $S = \{(X > 3) \wedge (Y = X)\}$, the initial cost was 2761 (281 U3 gates + 248 CX gates), which reduced by roughly 56\% to 1221 (141 U3 gates + 108 CX gates) after optimization. Table \ref{tab1} displays the qubit and gate count comparison for conjunctive formulae involving 2 variables for different number of clauses and bitstrings of different lengths. Further optimizations can also be done for a specific conjunctive formula $S$, taking into consideration the unused qubits in that particular case and using them as ancilla. It may be noted that since the control of quantum hardware is via analog pulses, pulse control techniques can show improvements over the gate-based compilation in processor error reduction. Such optimizations using Qiskit Pulse has been discussed in~\cite{Pulse}.

\section{Device run and noise analysis}

Due to the constraints on the quantum volume and number of qubits available in NISQ processors available over the cloud from IBM, our trials on real quantum processors were limited to satisfiability circuits for three-bit numbers. The circuit was run on \textbf{ibmq\_16\_melbourne} v2.3.23, belonging to the Canary Processors from IBM Quantum and having a quantum volume (QV)~\cite{QV} of 8. The conjunctive inequality, $S = \{(X < 5) \wedge (Y = 6)\}$ was checked for solutions. This implementation requires 15 qubits, same as the number of qubits available in the processor that was used. The optimizations mentioned in the previous section were done on the circuit and the optimization level of the transpiler was also set to 3 in all the executions.

\begin{figure}[!htbp] 
    \centering
  \subfloat[Results obtained on the \textbf{ibmq\_qasm\_simulator} with the probabilities of the 5 solutions ((6, 0), (6, 1), (6, 2), (6, 3) and (6, 4)) correctly amplified. $P_{\mbox{tot}} = 0.569.$\label{ideal_fig}]{%
       \includegraphics[width=0.99\linewidth]{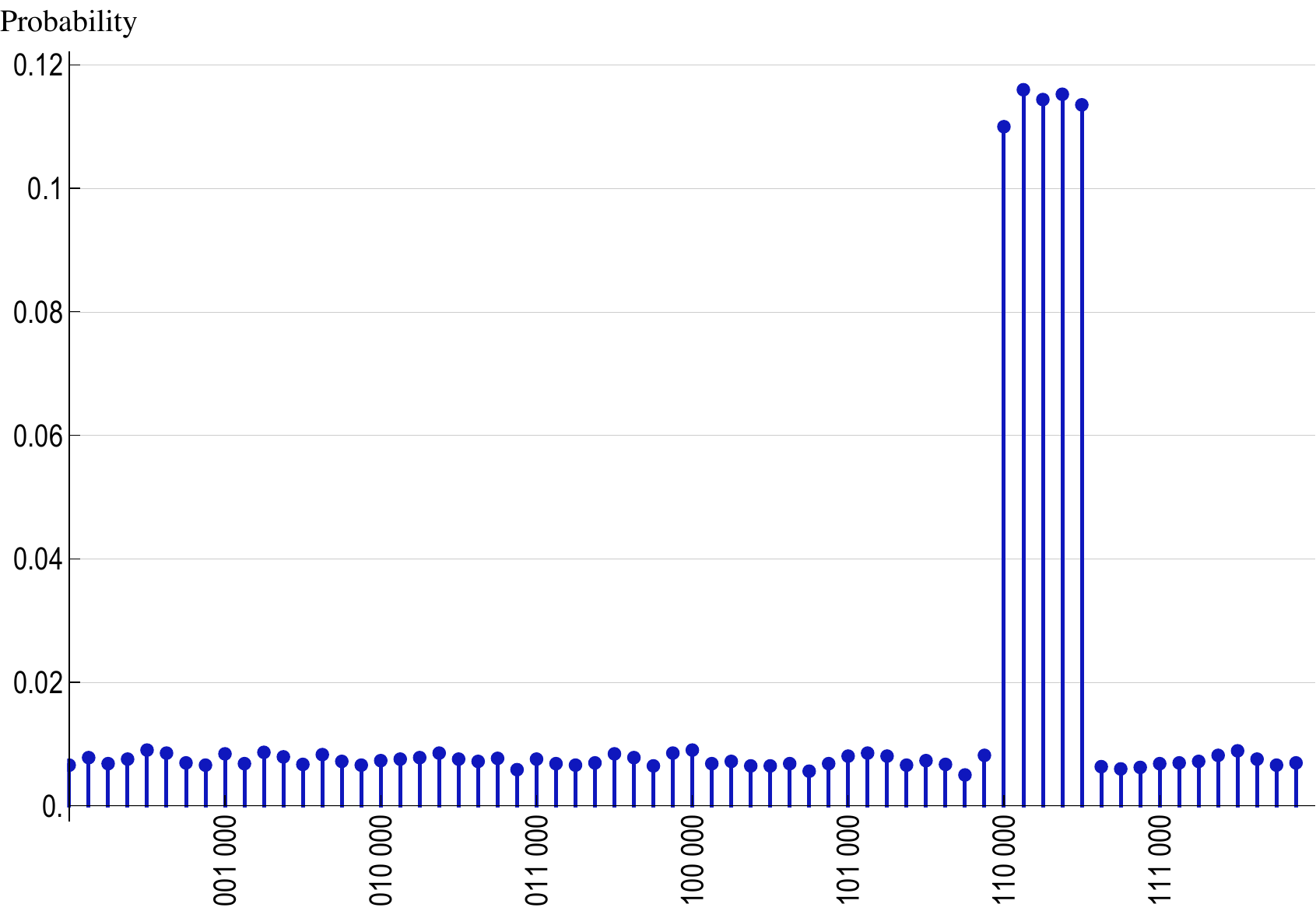}}
    \hfill
  \subfloat[The erroneous results (giving a random probability distribution) obtained on the \textbf{ibmq\_16\_melbourne} processor. $P_{\mbox{tot}} = 0.081.$\label{err_dev_fig}]{%
        \includegraphics[width=0.99\linewidth]{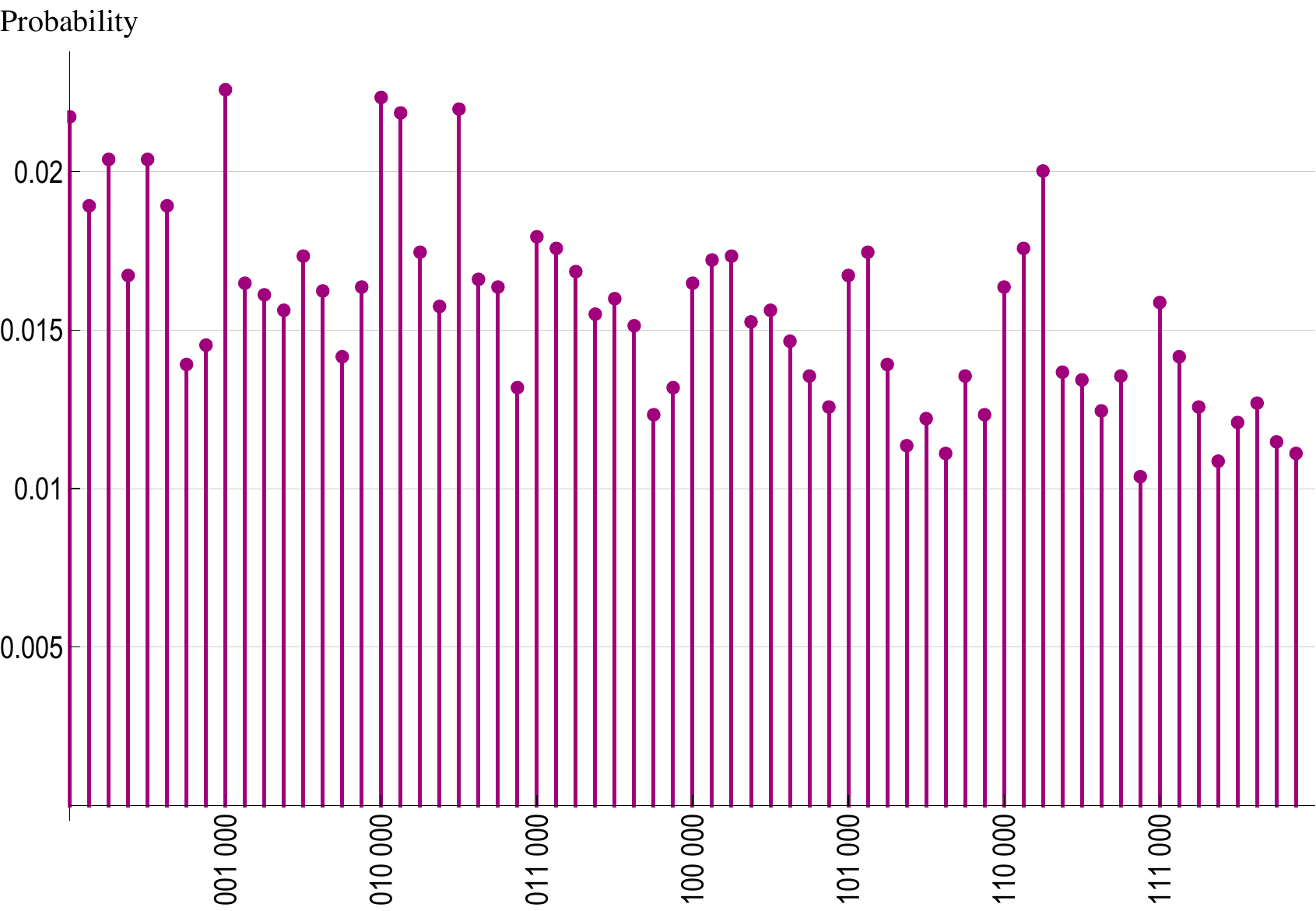}}
    \\
  \subfloat[Results obtained on the \textbf{ibmq\_qasm\_simulator} with the noise model, basis gate set and layout matching those of the \textbf{ibmq\_16\_melbourne} processor. $P_{\mbox{tot}} = 0.076.$\label{noise_sim_fig}]{%
        \includegraphics[width=0.99\linewidth]{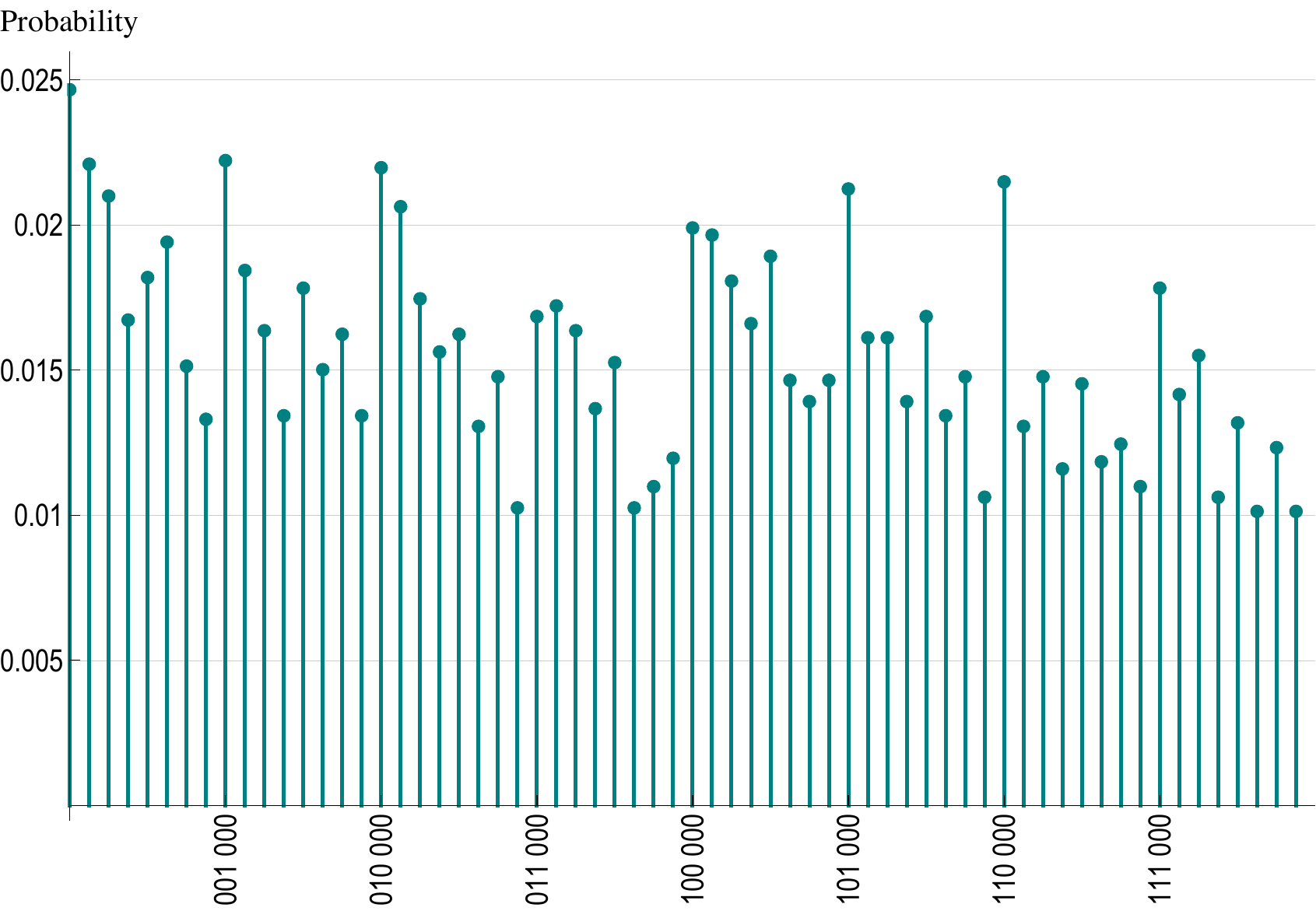}}
    \hfill
  \caption{Probability distributions on solving $S = \{(X < 5) \wedge (Y = 6)\}$. The circuit was run 8192 times in all the implementations.}
  \label{noise_quantify_fig_grp} 
\end{figure}

In Fig.\ref{ideal_fig}, the output obtained from running the circuit in the \textbf{ibmq\_qasm\_simulator} is shown. The circuit was unrolled into the general unitary gate and the controlled-X gate basis while implementing on the simulator. The correct solutions were obtained with high probability on the simulator. Fig.~\ref{err_dev_fig} shows the results of the same computation run on the \textbf{ibmq\_16\_melbourne} device. We see that the actual device fails to give the desired output and performs rather badly. In order to put the performance of the \textbf{ibmq\_16\_melbourne} device in context, we look at its Quantum Volume (QV) relative to the QV required to run a typical circuit we consider. The Quantum Volume (QV) of a device is the measure of the largest random circuit of equal width and depth that it can implement successfully. IBM's definition of quantum volume~\cite{QV} takes into account not only the number of qubits available for computation but also their connectivity, qubit and gate noise, etc. that are all factors limiting the computations that can be performed on a NISQ processor. To compute the QV of a device, the minimum of the number of qubits, $m$ and depth of the circuit, $d(m)$ that can be run with at least two-thirds probability of success given the noise in the circuit is computed first. This minimum is maximized over all possible proper and improper subsets of qubits in the device to obtain the logarithm of the QV. Qiskit provides for the necessary routines for estimating the QV of a given device. For a model circuit of $m$ qubits and depth $d(m)$, it has been estimated in~\cite{QV} that the computation fails with high probability when the model circuit volume satisfies the condition $md \approx 1/\epsilon_{\mathrm{eff}}$,  where $\epsilon_{\mathrm{eff}}$ is an estimate of the mean effective error probability per two-qubit gate. 

Two examples of conjunctive inequalities for 3-bit numbers were checked for solution. Out of these two examples, the case $S = \{(X < 5) \wedge (Y = 6)\}$ has a depth ($d$) of ~240 when unrolled into the most common device basis of the IBM Q systems (that includes the gates $R_Z$, $\sqrt{X}$, X and controlled-X). Since the circuit requires a minimum 15 qubits, in the absence of noise in the qubits or the gates and without connectivity restrictions (so that allowed circuit depth is unconstrained), the QV of the device required to run this circuit would be at least $2^{15} = 32,768$. The \textbf{ibmq\_16\_melbourne} device has 15 accessible qubits but it has a Quantum Volume of only 8 due to the significant noise levels present in the NISQ device. Hence the large circuit being implemented here using all 15 qubits is bound to fail, and this is seen to be so. In the execution on \textbf{ibmq\_16\_melbourne} device, the circuit was unrolled into the device's specific set of basis gates (I, $R_Z$, $\sqrt{X}$, X, controlled-X). The layout option of the Qiskit transpiler was also set to the `noise\_adaptive' option. Since noise constrains the QV of \textbf{ibmq\_16\_melbourne} device even if it has sufficient number of qubits to run our circuits successfully, we estimate in the following the threshold noise levels below which the processor can, in principle, execute the circuits.

To analyse the noisy result, the device noise parameters were included in the simulator, and the execution was done on this noisy simulator. Here, the same transpiled circuit as the one in the actual device run was used in the execution. A noise model having parameters matching those of the device backend, and the basis gates and corresponding coupling map for this configuration were used. These were automatically generated from the device backend properties using the Qiskit NoiseModel class. This is an approximate model comprising of the gate error probability and the gate length of each basis gate on each qubit, in addition to the relaxation time constants and the readout error probabilities of each qubit \cite{Qkit_doc}. The noisy result obtained (as shown in Fig.~\ref{noise_sim_fig}) confirms that the erroneous result on the actual processor can be attributed to the noise in the device.

Errors on the device were further studied to find a threshold for the noise below which the execution gives relatively better results. For this, a custom noise model was created with a thermal relaxation error and a depolarizing error channel. Using the device-specific compiled circuit as before, the circuit with error was simulated on the \textbf{ibmq\_qasm\_simulator}. The single-qubit gate error, comprising of the single-qubit depolarizing error followed by the single-qubit thermal relaxation error, was applied on all the single-qubit gates in the device basis (I, $R_Z$, $\sqrt{X}$ and X). The two-qubit gate error in the form of two-qubit depolarizing error followed by  single-qubit relaxation errors on the individual qubits coupled by the gate was applied on all the controlled-X gates \cite{Qkit_doc}. No other errors were considered for this custom noise model.

\begin{figure}[!htb] 
    \centering
  \subfloat[The result of the noisy simulation with the parameters of the thermal relaxation error channel matching those of \textbf{ibmq\_16\_melbourne} ($T_1 = 55.72 \mu$s, $T_2 = 60.51 \mu$s and $t$ = 928 ns), as on May 24, 2021; and those of the single and two qubit gate depolarizing errors as $\lambda_1 = 10^{-3}$ and $ \lambda_2 = 10^{-2}$, respectively. $P_{\mbox{tot}} = 0.080.$\label{noisy_iter1_fig}]{%
       \includegraphics[width=0.99\linewidth]{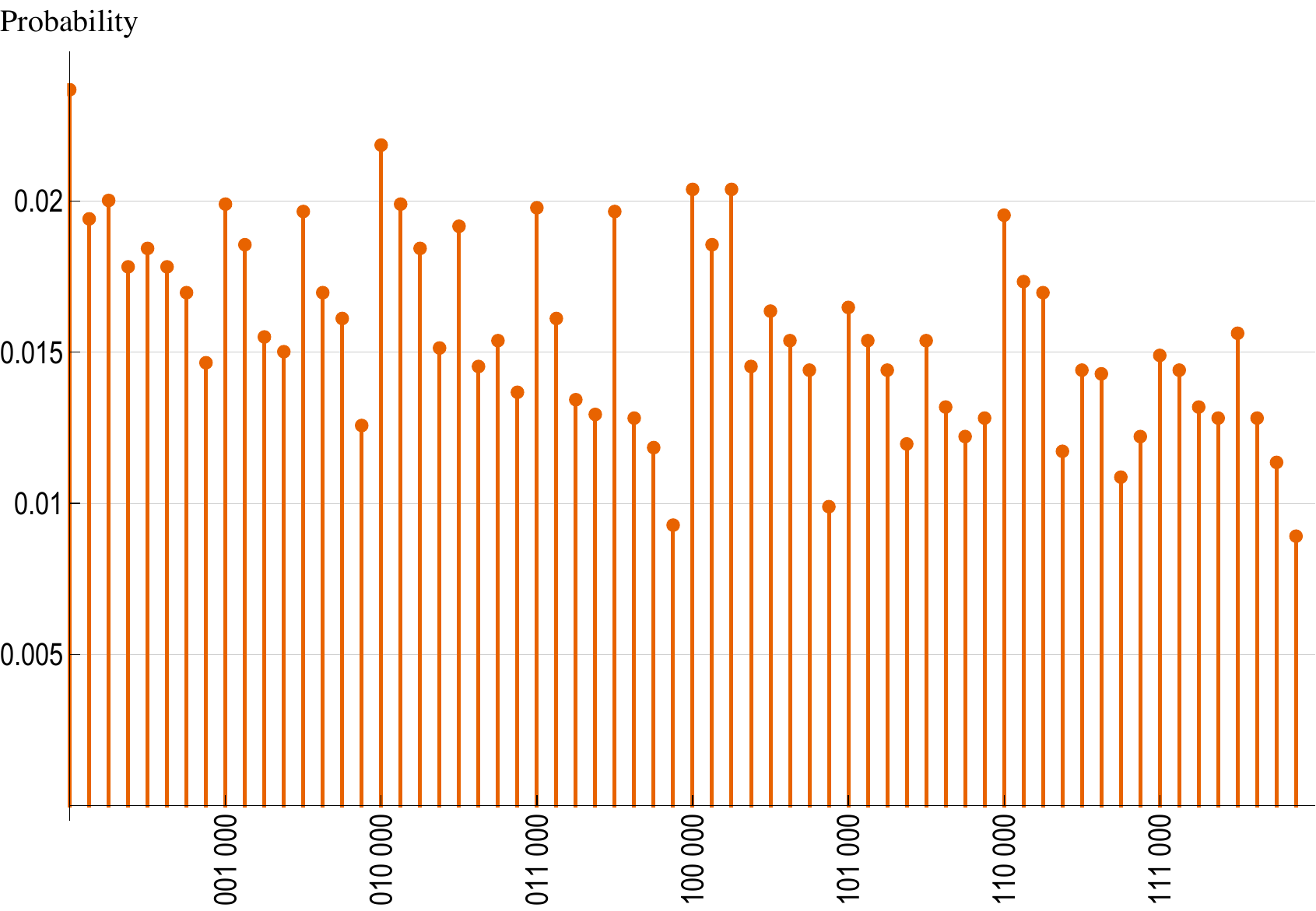}}
    \hfill
  \subfloat[The result of the noisy simulation with the parameters of the thermal relaxation error channel as $T_1 = 1155.72 \mu$s, $T_2 = 1160.51 \mu$s and $t$ = 598 ns; and those of the single and two qubit gate depolarizing errors as $\lambda_1 = 2.7 \times 10^{-4}$ and $\lambda_2 = 2.7 \times 10^{-3}$, respectively. $P_{\mbox{tot}} = 0.179.$\label{noisy_iter110_fig}]{%
        \includegraphics[width=0.99\linewidth]{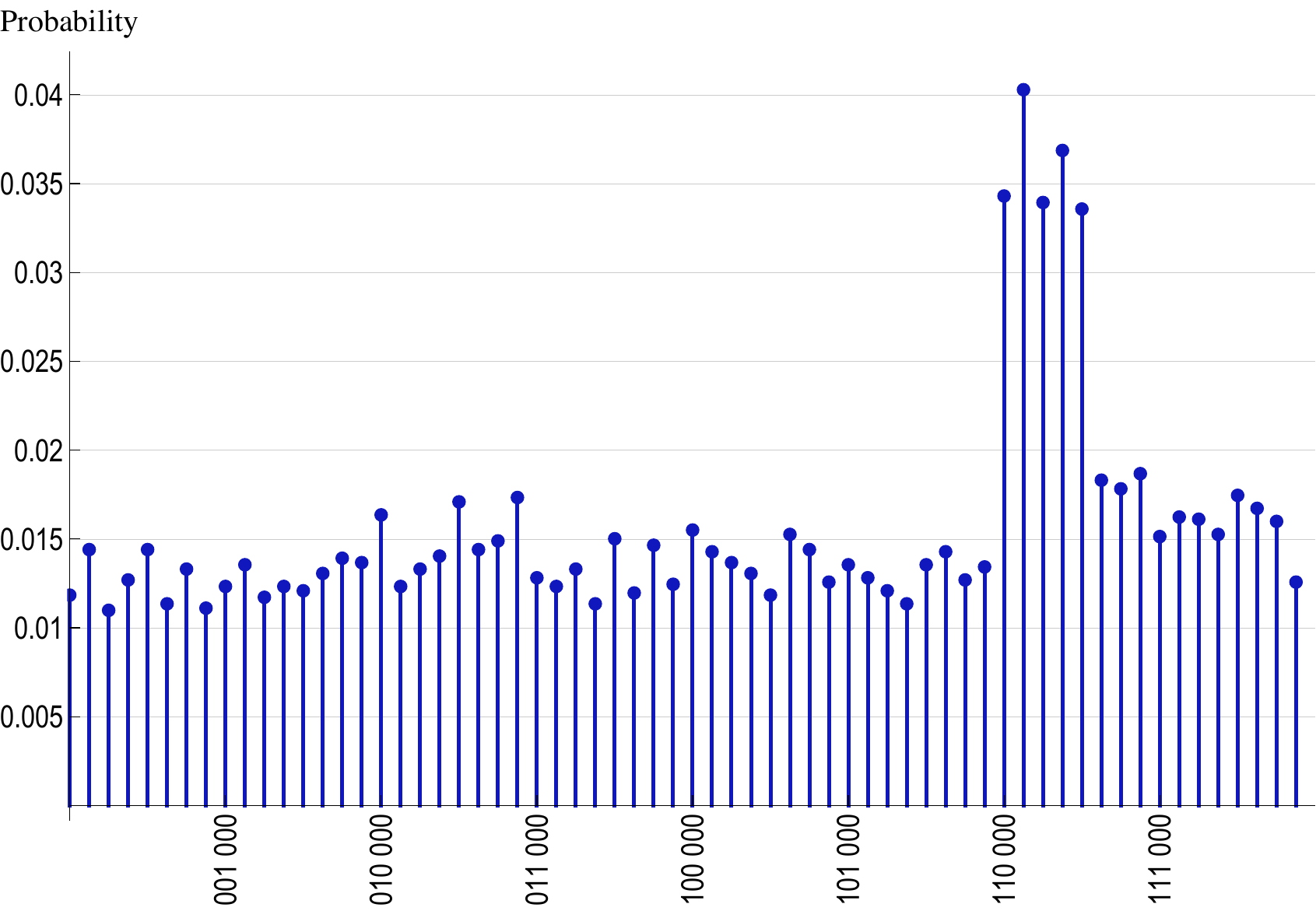}}
    \hfill
  \caption{Probability distributions on solving $S = \{(X < 5) \wedge (Y = 6)\}$ on the \textbf{ibmq\_qasm\_simulator} with the custom noise model. The circuit was run 8192 times in all the executions. Improvements in various noise parameters by factors ranging from 10 to 20 are required for \textbf{ibmq\_16\_melbourne} to start giving solutions for the three bit problems we have considered.}
  \label{noise_iter_fig_grp} 
\end{figure}

    \begin{figure}[ht!]
    \includegraphics[width=0.99\linewidth]{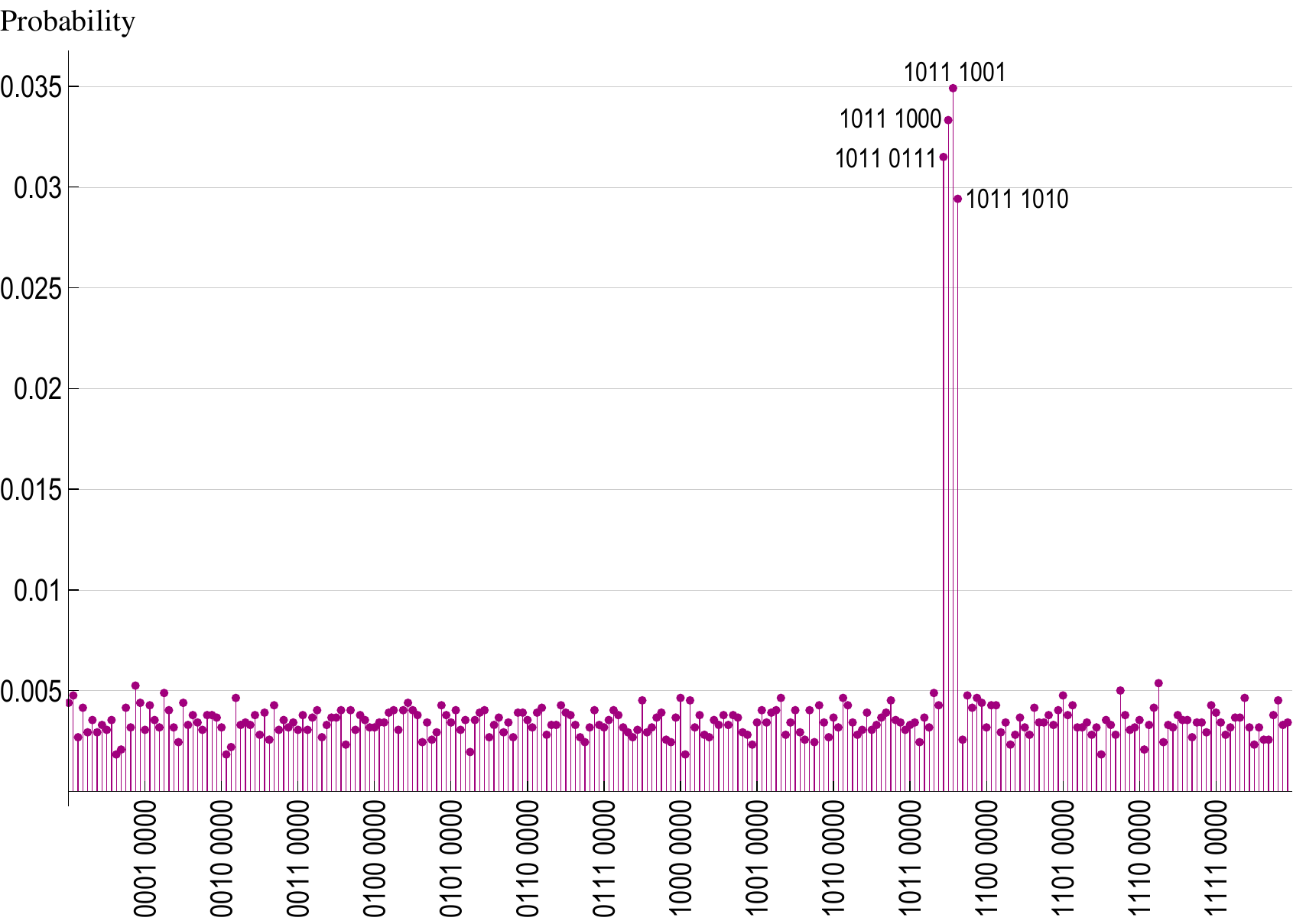}
    \caption{$S = \{(X < 14) \wedge (X > 6) \wedge (Y = 11) \wedge (X < Y)\}$ with the probabilities of the 4 solutions ((11, 7), (11, 8), (11, 9) and (11, 10)) amplified. $P_{\mbox{tot}} = 0.128.$ \label{out1_fig}}
    \end{figure}
    
    \begin{figure}[ht!]
    \includegraphics[width=0.99\linewidth]{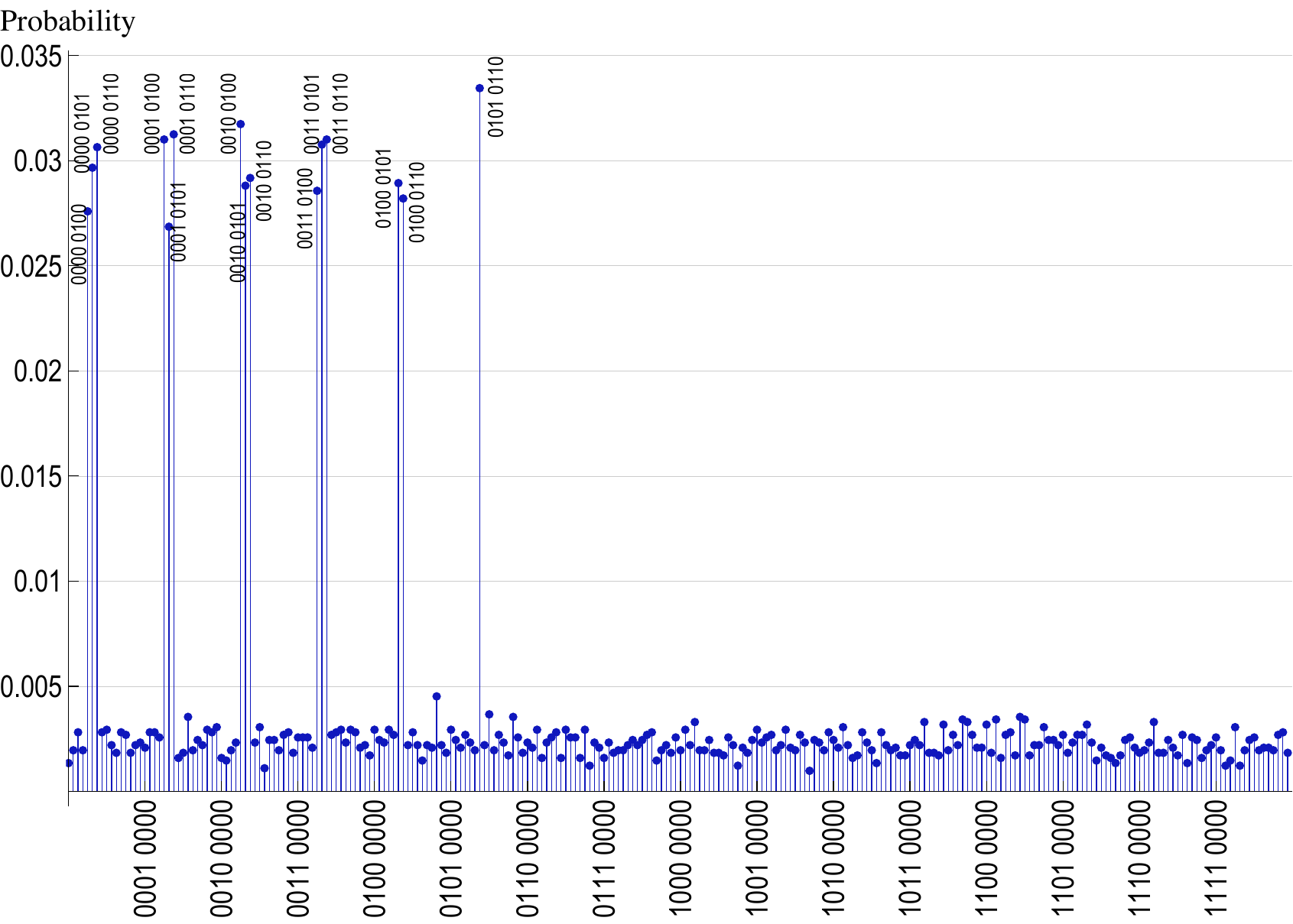}
    \caption{$S = \{(X < 7) \wedge (X > 3) \wedge (Y < X)\}$ with the probabilities of the 15 solutions ((0, 4), (0, 5), (0, 6), (1, 4), (1, 5), (1, 6), (2, 4), (2, 5), (2, 6), (3, 4), (3, 5), (3, 6), (4, 5), (4, 6) and (5, 6)) amplified. $P_{\mbox{tot}} = 0.449.$ \label{out2_fig}}
    \end{figure}
    
    \begin{figure}[ht!]
    \includegraphics[width=0.99\linewidth]{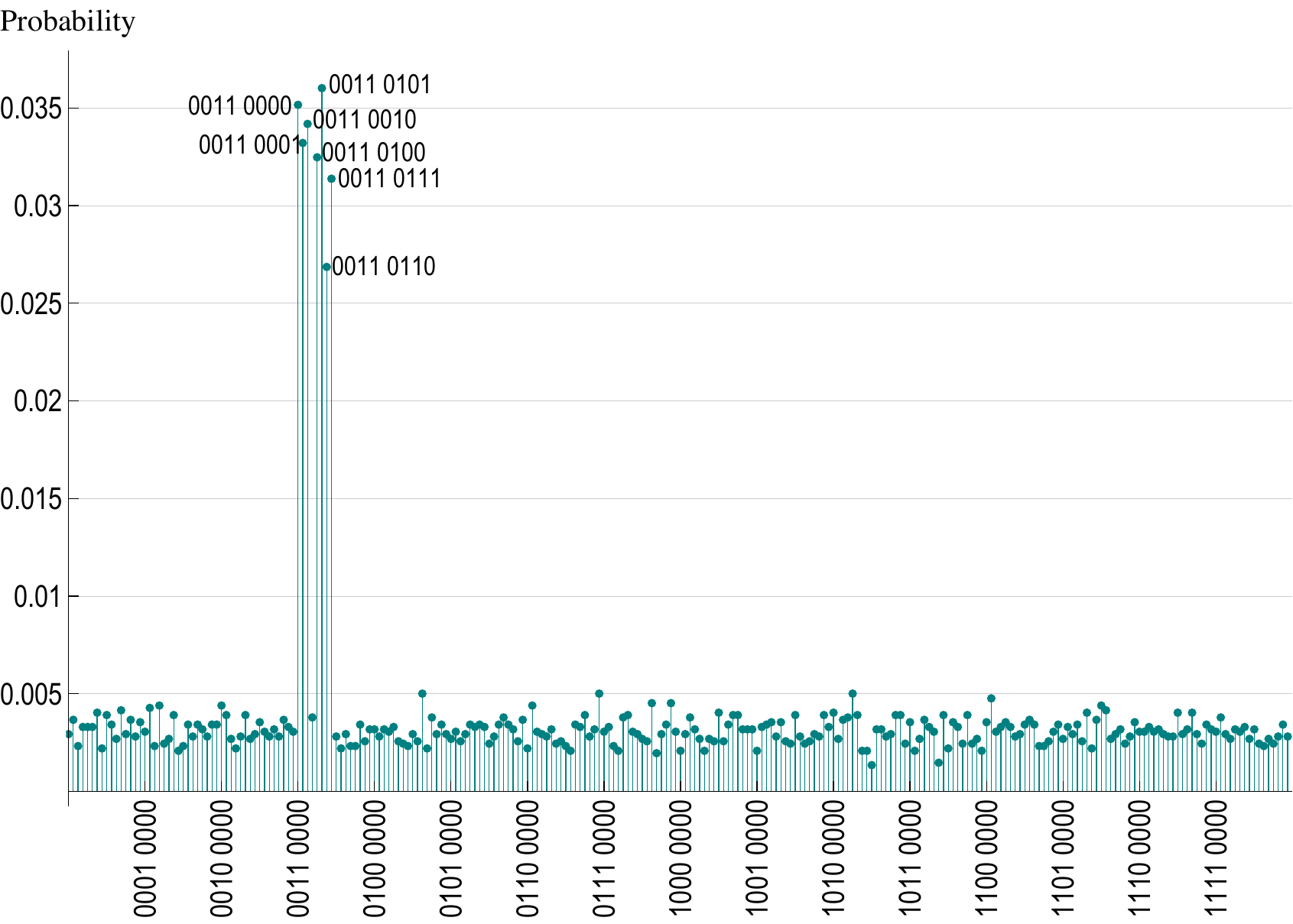}
    \caption{$S = \{(X < 8) \wedge (Y = 3) \wedge (X \neq Y)\}$ with the probabilities of the 7 solutions ((3, 0), (3, 1), (3, 2), (3, 4), (3, 5), (3, 6) and (3, 7)) amplified. $P_{\mbox{tot}} = 0.228.$ \label{out3_fig}}
    \end{figure}
    
    \begin{figure}[ht!]
    \includegraphics[width=0.99\linewidth]{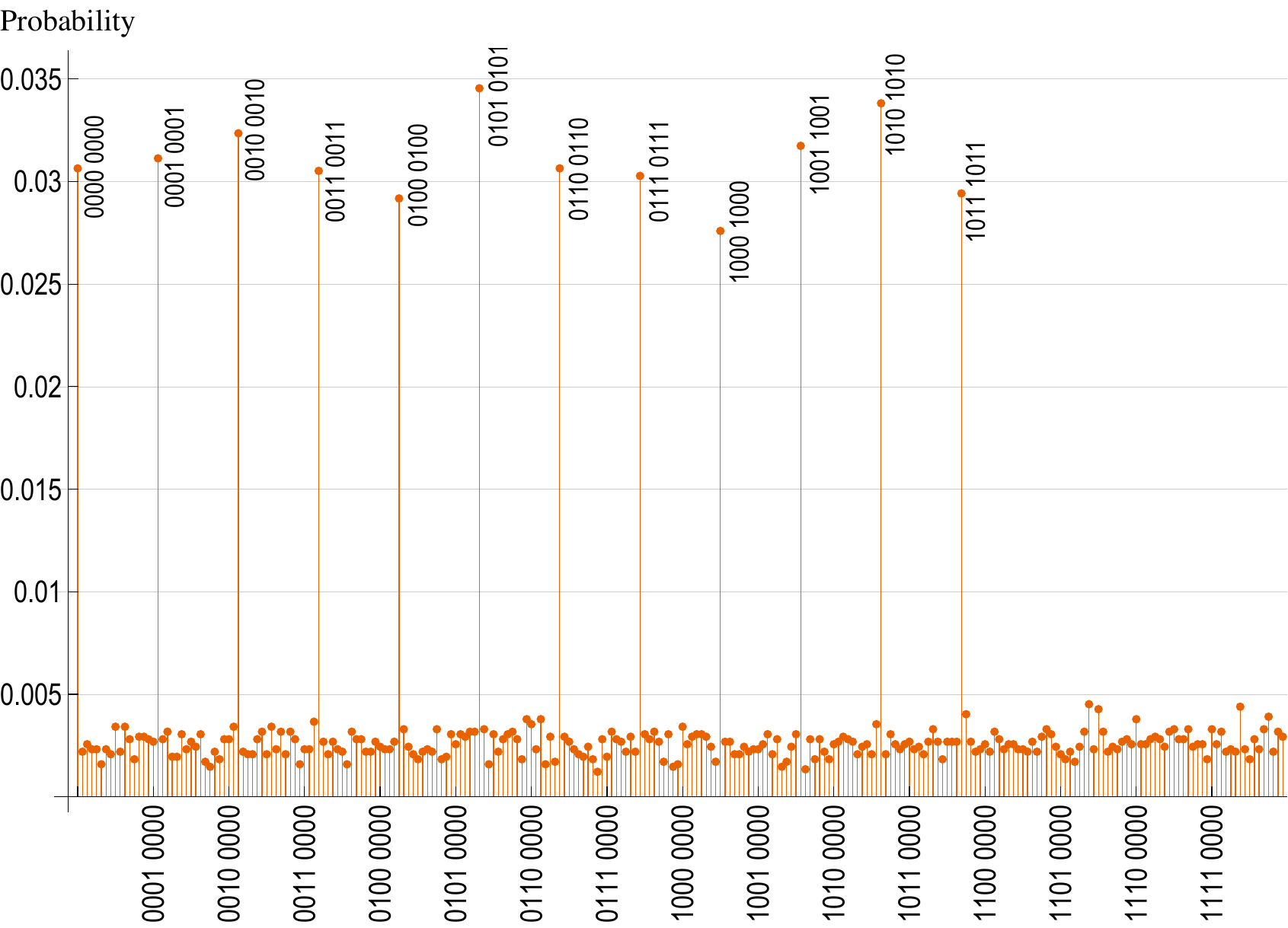}
    \caption{$S = \{(X < 12) \wedge (Y = X)\}$ with the probabilities of the 12 solutions ((0, 0), (1, 1), (2, 2), (3, 3), (4, 4), (5, 5), (6, 6), (7, 7), (8, 8), (9, 9), (10, 10) and (11, 11)) amplified. $P_{\mbox{tot}} = 0.373.$ \label{out4_fig}}
    \end{figure}
    
The parameters for the single-qubit thermal relaxation error were chosen to match those of the device (\textbf{ibmq\_16\_melbourne}). The average values of the thermal relaxation time constant $(T_1)$ and the dephasing time constant $(T_2)$ were set to be 55.72 $\mu$s and 60.51 $\mu$s respectively. The gate time ($t$) for relaxation error was set to 928 ns. These parameter values are as obtained from the calibration information of the device from~\cite{IBMQ} as on May 24, 2021.  The executions on the real device in Fig.~\ref{noise_quantify_fig_grp} were also performed on the same date, so that the comparisons done would be meaningful. The depolarizing error channel parameters were chosen arbitrarily to be $10^{-3}$ and $10^{-2}$ respectively for the single and two-qubit gates $(\lambda_1$ and $\lambda_2$ respectively).  With the initially chosen parameter values, the obtained result was noisy, as seen in Fig.~\ref{noisy_iter1_fig}, with significant errors involved. The noise parameters were progressively changed by increasing $T_1$ and $T_2$ by 10 $\mu$s, reducing $t$ by 3 ns, and reducing $\lambda_1$ and $\lambda_2$ by $6.6 \times 10^{-6}$ and $6.6 \times 10^{-5}$ respectively, in each step. The results were seen to get better and satisfactory results were obtained with high probability for $T_1 = 1155.72$ $\mu$s, $T_2 = 1160.51$ $\mu$s, $t$ = 598 ns, $\lambda_1 = 2.7 \times 10^{-4}$ and $\lambda_2 = 2.7 \times 10^{-3}$ as shown in Fig.~\ref{noisy_iter110_fig}. The total success probability, $P_{\mbox{tot}}$, in this case is not as high as that for the ideal simulation. This is because of the noise present, which increases the probabilities of the other states as well. However, it is still good enough to clearly determine the solutions, when repeated several times. Another conjunctive formula, $S = \{(X > 3) \wedge (Y = X)\}$ was also checked for solution with these parameters and again the desired results were obtained at these noise levels.
 
The simulation with noise is an approximate study to emphasize the fact that better results can be obtained by reducing the noise. Since all types of errors were not considered in this model, the actual device run may require different ranges of values of these parameters. The values of the parameters obtained here is just one example giving reasonable results. In reality, the ease of mitigating some kinds of errors can be leveraged to obtain better results by mainly controlling such errors. The circuits can also be altered with different approximation mechanisms to obtain sufficiently accurate results on these NISQ devices, instead of targeting a significant improvement in the hardware calibration alone.

Figures \ref{out1_fig} to \ref{out4_fig} show some more examples solved using the algorithm implementation on the \textbf{ibmq\_qasm\_simulator} with the circuits optimized as described in Sec.~\ref{optimization}. The solution pairs are ordered as $(Y, X)$. All the circuits were run 8192 times to obtain the final probability distributions. The expected probability distributions were obtained in one Grover iteration itself in all these cases.

\section{Discussion and Conclusion}

Motivated by the importance of finding solutions for conjunctive formulae in classical computation problems of practical importance, we explored the solution of such formulae in NISQ devices. Specifically, we looked at conjunctive formulae on the integer domain. Casting the problem as a database search, we employed Grover's algorithm for finding the solutions. The individual component circuits for various comparison operators were worked out. We showed how they can be combined to form the oracle for the Grover's algorithm corresponding to the search of solutions for various conjunctive formulae. Our approach not only checks for the satisfiability of conjunctive formulae but seeks out the variable values that satisfy them as well. Combining the oracle with the diffusion operation, we constructed the full Grover iteration. We also showed how the cost of implementing the circuits can be reduced by optimizing the gates. 

Worked out examples showed that for conjunctive formulae involving two variables and constants that are all 3- or 4-bit numbers, our implementation of Grover's algorithm can find the solutions within one iteration when running on the noise-free classical simulators of IBM Q devices. On real devices, we see that while the quantum volume is sufficient to run some of the simpler cases, the noise levels are still too high to obtain usable results. We verify that the noise is indeed the main impediment in implementation on NISQ devices by running the simulation with added noise and seeing the degradation in performance. We are able to establish noise thresholds below which actual devices can lead to good results, thereby setting targets for noise reduction and augmenting the case for quantum error correction. 

Our implementation provides the framework for solving an important class of problems in the short term on real devices. With the quantum volume of available devices expected to improve rapidly with noise levels going down and further gate optimizations possible, successful execution and identification of the  solutions of large classes of conjunctive formulae on the integer domain will come within reach. Extension of our approach to constraint satisfiability problems on the real domain is also quite straightforward. Multiple applications of the available gates like $H$ and $T$ can approximate the required rotation gates~\cite{qkit} to facilitate such an extension.
	
\begin{acknowledgements}
This work was supported in part by the Science and Engineering Research Board of the Department of Science and Technology, Government of India through grant No. EMR/2016/007221 and in part by the QuEST program of the Department of Science and Technology, through project No. Q113 under Theme 4. We acknowledge the use of IBM Quantum services for this work. The views expressed are those of the authors, and do not reflect the official policy or position of IBM or the IBM Quantum team. G.~V. thanks all the mentors and participants of the IBM Quantum Challenge Fall 2020 for useful discussions.

\end{acknowledgements}
	
\bibliography{final_ref_arxiv}
\end{document}